\documentclass[manuscript,screen,review=false]{jair}

\usepackage{array}
\usepackage{tabularx}
\usepackage{longtable}
\usepackage{algorithm}
\usepackage[noend]{algpseudocode}
\usepackage{placeins}
\usepackage{flafter}
\usepackage{needspace}
\usepackage{pict2e}
\usepackage{tikz}
\usepackage{pgfplots}
\usepackage{pgfplotstable}
\usetikzlibrary{arrows.meta,positioning,calc,patterns.meta}
\usepgfplotslibrary{statistics}
\pgfplotsset{compat=1.18}
\tikzset{
  fig text/.style={font=\sffamily\fontsize{8}{11.5}\selectfont,text=black,inner sep=0pt},
  fig heading/.style={fig text,font=\sffamily\bfseries\fontsize{8.5}{10}\selectfont,anchor=west},
  fig box/.style={fig text,draw=black!70,line width=.45pt,fill=white,
    align=center,inner sep=2pt,minimum height=8mm},
  fig arrow/.style={-{Stealth[length=1.7mm,width=1.2mm]},draw=black!75,line width=.5pt},
  fig link/.style={draw=black!65,line width=.45pt},
  fig tinybox/.style={fig text,draw=black!65,line width=.4pt,inner xsep=2pt,inner ysep=4pt},
}
\pgfplotsset{figure axis/.style={
  font=\sffamily\fontsize{8}{9.5}\selectfont,
  tick label style={font=\sffamily\fontsize{8}{9.5}\selectfont},
  xticklabel style={yshift=-1mm},
  label style={font=\sffamily\fontsize{8}{9.5}\selectfont},
  xlabel style={yshift=-3mm},
  axis lines*=left,axis line style={draw=black!65,line width=.45pt},
  tick style={draw=black!65,line width=.4pt},
  major tick length=1mm,minor tick length=0pt,
  scaled ticks=false,clip=false,
}}

\makeatletter
\newdimen\FigLeft\newdimen\FigBottom\newdimen\FigRight\newdimen\FigTop
\newdimen\FigLabelX\newdimen\FigLabelY
\newcommand{\ReportFigurePanel}[5]{%
  \pgfextractx{\FigLeft}{\pgfpointanchor{#3}{center}}%
  \pgfextracty{\FigBottom}{\pgfpointanchor{#3}{center}}%
  \pgfextractx{\FigRight}{\pgfpointanchor{#4}{center}}%
  \pgfextracty{\FigTop}{\pgfpointanchor{#4}{center}}%
  \pgfextractx{\FigLabelX}{\pgfpointanchor{#5}{west}}%
  \pgfextracty{\FigLabelY}{\pgfpointanchor{#5}{west}}%
  \typeout{FIGPANEL|#1|#2|\strip@pt\FigLeft|\strip@pt\FigBottom|\strip@pt\FigRight|\strip@pt\FigTop|\strip@pt\FigLabelX|\strip@pt\FigLabelY}%
}
\makeatother

\newcolumntype{L}[1]{>{\raggedright\arraybackslash}p{#1}}
\newlength{\settingcolumnwidth}
\newlength{\qualitycolumnwidth}
\newlength{\studiescolumnwidth}
\newlength{\directcolumnwidth}
\newlength{\basecolumnwidth}
\newcommand{\tablecheck}{%
  \begingroup
  \setlength{\unitlength}{1pt}%
  \raisebox{0.25ex}{\begin{picture}(7,5)
    \linethickness{0.65pt}%
    \put(0,2){\line(1,-1){2}}%
    \put(2,0){\line(1,1){5}}%
  \end{picture}}%
  \endgroup}
\theoremstyle{acmplain}
\newtheorem{theorem}{Theorem}[section]
\newtheorem{lemma}[theorem]{Lemma}
\newtheorem{proposition}[theorem]{Proposition}
\theoremstyle{acmdefinition}
\newtheorem{definition}[theorem]{Definition}
\theoremstyle{acmplain}

\setcopyright{cc}
\acmDOI{}
\JAIRAE{}
\JAIRTrack{}
\acmVolume{}
\acmArticle{}
\acmYear{2026}
\makeatletter
\let\JAIROriginalFormatDOI\@formatdoi
\renewcommand{\@formatdoi}[1]{%
  \ifx\@acmDOI\@empty\else\JAIROriginalFormatDOI{#1}\fi}
\makeatother
\AtBeginDocument{%
  \fancyfoot[RO,LE]{\footnotesize Manuscript submitted to the Journal of Artificial Intelligence Research.}%
  \fancypagestyle{firstpagestyle}{%
    \fancyhf{}%
    \fancyfoot[RO,LE]{\footnotesize Manuscript submitted to the Journal of Artificial Intelligence Research.}%
  }%
}

\usepackage[
  datamodel=acmdatamodel,
  style=acmauthoryear,
  backend=biber,
  giveninits=true,
  uniquename=init,
  uniquelist=false,
  maxcitenames=2,
  mincitenames=1
]{biblatex}

\begin{document}

\newcommand{\PriorityQualityN}{%
  68%
}

\newcommand{\MainQualityN}{%
  84%
}

\newcommand{\MainQualityMedianRows}{%
  \textsc{BG-d$\infty$} & \textbf{1.5} & 2.5 & 2.5 & \textbf{5.0} & 0.518 \\%
  \textsc{IR} & 35.0 & \textbf{1.0} & 2.0 & 45.5 & 0.493 \\%
  \textsc{IR-IM-IS} & 4.0 & \textbf{1.0} & \textbf{1.0} & 6.5 & \textbf{0.347} \\%
}

\newcommand{\MainQualityMaximumLower}{%
  11%
}

\newcommand{\MainQualityTotalLower}{%
  56%
}

\newcommand{\StressQualityN}{%
  47%
}

\newcommand{\StressQualityTotalLower}{%
  22%
}

\newcommand{\StressHighQualityN}{%
  21%
}

\newcommand{\StressHighQualityTotalLower}{%
  20%
}

\newcommand{\MatchedFormulaRows}{%
  \multicolumn{4}{@{}l}{\textit{All-126 ($n=126$)}} \\%
  \textsc{Direct} & 170{,}517 & 93{,}183.5 & 93{,}609.5 \\%
  \textsc{Compact-Base} \textbf{(ours)} & 81{,}521.5 & 72{,}487 & 72{,}612.5 \\%
  \textsc{Compact-Objective} \textbf{(ours)} & \textbf{66{,}318} & \textbf{67{,}255.5} & \textbf{67{,}381} \\%
  \addlinespace%
  \multicolumn{4}{@{}l}{\textit{Stress ($n=60$)}} \\%
  \textsc{Direct} & 176{,}987 & 102{,}284 & 102{,}704 \\%
  \textsc{Compact-Base} \textbf{(ours)} & 80{,}066.5 & 70{,}995.5 & 71{,}119.5 \\%
  \textsc{Compact-Objective} \textbf{(ours)} & \textbf{65{,}562.5} & \textbf{66{,}475.5} & \textbf{66{,}599.5} \\%
  \addlinespace%
  \multicolumn{4}{@{}l}{\textit{Stress-high ($n=40$)}} \\%
  \textsc{Direct} & 182{,}322 & 107{,}596 & 108{,}076 \\%
  \textsc{Compact-Base} \textbf{(ours)} & 73{,}728 & 64{,}861.5 & 64{,}985 \\%
  \textsc{Compact-Objective} \textbf{(ours)} & \textbf{60{,}638} & \textbf{61{,}553.5} & \textbf{61{,}677} \\%
}

\newcommand{\MatchedRuntimePrimaryRows}{%
  \multicolumn{5}{@{}l}{\textit{All-126 ($n=126$)}} \\%
  \textsc{BG-d2} & \textsc{Direct} & 87/39/0 & 9.3 & 144.8 \\%
   & \textsc{Compact-Base} \textbf{(ours)} & 87/39/0 & \textbf{2.9} & 73.7 \\%
   & \textsc{Compact-Objective} \textbf{(ours)} & 87/39/0 & 3.9 & \textbf{66.6} \\%
  \textsc{IR} & \textsc{Direct} & 89/37/0 & 1.8 & 74.8 \\%
   & \textsc{Compact-Base} \textbf{(ours)} & 89/37/0 & \textbf{0.8} & 66.9 \\%
   & \textsc{Compact-Objective} \textbf{(ours)} & 89/37/0 & \textbf{0.8} & \textbf{66.2} \\%
  \textsc{IR-IM-IS} & \textsc{Direct} & \textbf{84}/37/5 & \textbf{637.0} & 78.3 \\%
   & \textsc{Compact-Base} \textbf{(ours)} & 83/37/6 & 722.4 & 69.2 \\%
   & \textsc{Compact-Objective} \textbf{(ours)} & \textbf{84}/37/5 & 649.1 & \textbf{67.0} \\%
  \addlinespace%
  \multicolumn{5}{@{}l}{\textit{Stress ($n=60$)}} \\%
  \textsc{BG-d2} & \textsc{Direct} & \textbf{56}/0/4 & 2{,}186.1 & 149.3 \\%
   & \textsc{Compact-Base} \textbf{(ours)} & \textbf{56}/0/4 & \textbf{1{,}199.4} & 80.3 \\%
   & \textsc{Compact-Objective} \textbf{(ours)} & 55/0/5 & 1{,}779.6 & \textbf{71.3} \\%
  \textsc{IR} & \textsc{Direct} & 60/0/0 & 7.6 & 87.8 \\%
   & \textsc{Compact-Base} \textbf{(ours)} & 60/0/0 & \textbf{3.9} & \textbf{69.9} \\%
   & \textsc{Compact-Objective} \textbf{(ours)} & 60/0/0 & 11.1 & 71.5 \\%
  \textsc{IR-IM-IS} & \textsc{Direct} & \textbf{48}/0/12 & \textbf{3{,}014.1} & 89.1 \\%
   & \textsc{Compact-Base} \textbf{(ours)} & 47/0/13 & 3{,}156.5 & 74.6 \\%
   & \textsc{Compact-Objective} \textbf{(ours)} & 47/0/13 & 3{,}189.3 & \textbf{73.3} \\%
  \addlinespace%
  \multicolumn{5}{@{}l}{\textit{Stress-high ($n=40$)}} \\%
  \textsc{BG-d2} & \textsc{Direct} & 28/0/12 & 5{,}239.1 & 152.2 \\%
   & \textsc{Compact-Base} \textbf{(ours)} & \textbf{29}/0/11 & \textbf{4{,}321.3} & 78.7 \\%
   & \textsc{Compact-Objective} \textbf{(ours)} & 27/0/13 & 5{,}011.1 & \textbf{71.7} \\%
  \textsc{IR} & \textsc{Direct} & 40/0/0 & 4.1 & 89.2 \\%
   & \textsc{Compact-Base} \textbf{(ours)} & 40/0/0 & \textbf{2.1} & \textbf{68.3} \\%
   & \textsc{Compact-Objective} \textbf{(ours)} & 40/0/0 & 3.7 & \textbf{68.3} \\%
  \textsc{IR-IM-IS} & \textsc{Direct} & 20/0/20 & 7{,}525.8 & 158.4 \\%
   & \textsc{Compact-Base} \textbf{(ours)} & 20/0/20 & 7{,}492.2 & 163.8 \\%
   & \textsc{Compact-Objective} \textbf{(ours)} & \textbf{21}/0/19 & \textbf{7{,}436.9} & \textbf{142.7} \\%
}

\newcommand{\MatchedFormulaDetailRows}{%
  All-126 & \textsc{BG-d2} & \textsc{Direct} & 79{,}877 & 170{,}517 & 424{,}056 \\%
   &  & \textsc{Compact-Base} \textbf{(ours)} & 31{,}250.5 & 81{,}521.5 & 197{,}520.5 \\%
   &  & \textsc{Compact-Objective} \textbf{(ours)} & 27{,}736 & 66{,}318 & 153{,}675 \\%
  \addlinespace[1pt]%
   & \textsc{IR} & \textsc{Direct} & 27{,}859 & 93{,}183.5 & 213{,}157.5 \\%
   &  & \textsc{Compact-Base} \textbf{(ours)} & 26{,}727.5 & 72{,}487 & 167{,}045.5 \\%
   &  & \textsc{Compact-Objective} \textbf{(ours)} & 27{,}750 & 67{,}255.5 & 156{,}592 \\%
  \addlinespace[1pt]%
   & \textsc{IR-IM-IS} & \textsc{Direct} & 27{,}859 & 93{,}609.5 & 213{,}577.5 \\%
   &  & \textsc{Compact-Base} \textbf{(ours)} & 26{,}727.5 & 72{,}612.5 & 167{,}171 \\%
   &  & \textsc{Compact-Objective} \textbf{(ours)} & 27{,}750 & 67{,}381 & 156{,}717.5 \\%
  \addlinespace%
  Stress & \textsc{BG-d2} & \textsc{Direct} & 80{,}847 & 176{,}987 & 436{,}695 \\%
   &  & \textsc{Compact-Base} \textbf{(ours)} & 31{,}339.5 & 80{,}066.5 & 194{,}137 \\%
   &  & \textsc{Compact-Objective} \textbf{(ours)} & 27{,}556.5 & 65{,}562.5 & 151{,}679 \\%
  \addlinespace[1pt]%
   & \textsc{IR} & \textsc{Direct} & 27{,}915.5 & 102{,}284 & 231{,}278 \\%
   &  & \textsc{Compact-Base} \textbf{(ours)} & 26{,}802 & 70{,}995.5 & 163{,}552 \\%
   &  & \textsc{Compact-Objective} \textbf{(ours)} & 27{,}569.5 & 66{,}475.5 & 154{,}512 \\%
  \addlinespace[1pt]%
   & \textsc{IR-IM-IS} & \textsc{Direct} & 27{,}915.5 & 102{,}704 & 231{,}698 \\%
   &  & \textsc{Compact-Base} \textbf{(ours)} & 26{,}802 & 71{,}119.5 & 163{,}676 \\%
   &  & \textsc{Compact-Objective} \textbf{(ours)} & 27{,}569.5 & 66{,}599.5 & 154{,}636 \\%
  \addlinespace%
  Stress-high & \textsc{BG-d2} & \textsc{Direct} & 80{,}847 & 182{,}322 & 447{,}365 \\%
   &  & \textsc{Compact-Base} \textbf{(ours)} & 30{,}027.5 & 73{,}728 & 179{,}370.5 \\%
   &  & \textsc{Compact-Objective} \textbf{(ours)} & 26{,}015 & 60{,}638 & 140{,}049.5 \\%
  \addlinespace[1pt]%
   & \textsc{IR} & \textsc{Direct} & 27{,}915.5 & 107{,}596 & 241{,}882 \\%
   &  & \textsc{Compact-Base} \textbf{(ours)} & 25{,}537.5 & 64{,}861.5 & 149{,}516 \\%
   &  & \textsc{Compact-Objective} \textbf{(ours)} & 26{,}028 & 61{,}553.5 & 142{,}898.5 \\%
  \addlinespace[1pt]%
   & \textsc{IR-IM-IS} & \textsc{Direct} & 27{,}915.5 & 108{,}076 & 242{,}362 \\%
   &  & \textsc{Compact-Base} \textbf{(ours)} & 25{,}537.5 & 64{,}985 & 149{,}639.5 \\%
   &  & \textsc{Compact-Objective} \textbf{(ours)} & 26{,}028 & 61{,}677 & 143{,}022 \\%
}

\newcommand{\MatchedRuntimeAllRows}{%
  \multicolumn{8}{@{}l}{\textit{All-126 ($n=126$)}} \\%
  \textsc{BG-d2} & \textsc{Direct} & \textbf{87} & 39 & \textbf{0} & 1.740 & 9.254 & 144.81 \\%
   & \textsc{Compact-Base} \textbf{(ours)} & \textbf{87} & 39 & \textbf{0} & 0.716 & \textbf{2.948} & 73.73 \\%
   & \textsc{Compact-Objective} \textbf{(ours)} & \textbf{87} & 39 & \textbf{0} & \textbf{0.605} & 3.949 & \textbf{66.55} \\%
  \addlinespace[1pt]%
  \textsc{IR} & \textsc{Direct} & \textbf{89} & 37 & \textbf{0} & \textbf{0.494} & 1.757 & 74.77 \\%
   & \textsc{Compact-Base} \textbf{(ours)} & \textbf{89} & 37 & \textbf{0} & 0.497 & 0.840 & 66.91 \\%
   & \textsc{Compact-Objective} \textbf{(ours)} & \textbf{89} & 37 & \textbf{0} & 0.522 & \textbf{0.812} & \textbf{66.18} \\%
  \addlinespace[1pt]%
  \textsc{IR-IM-IS} & \textsc{Direct} & \textbf{84} & 37 & \textbf{5} & 1.201 & \textbf{637.033} & 78.34 \\%
   & \textsc{Compact-Base} \textbf{(ours)} & 83 & 37 & 6 & \textbf{0.573} & 722.384 & 69.25 \\%
   & \textsc{Compact-Objective} \textbf{(ours)} & \textbf{84} & 37 & \textbf{5} & 0.605 & 649.128 & \textbf{67.05} \\%
}

\newcommand{\MatchedRuntimeStressRows}{%
  \multicolumn{8}{@{}l}{\textit{Stress ($n=60$)}} \\%
  \textsc{BG-d2} & \textsc{Direct} & \textbf{56} & 0 & \textbf{4} & 10.143 & 2{,}186.134 & 149.33 \\%
   & \textsc{Compact-Base} \textbf{(ours)} & \textbf{56} & 0 & \textbf{4} & \textbf{2.827} & \textbf{1{,}199.383} & 80.31 \\%
   & \textsc{Compact-Objective} \textbf{(ours)} & 55 & 0 & 5 & 3.801 & 1{,}779.560 & \textbf{71.32} \\%
  \addlinespace[1pt]%
  \textsc{IR} & \textsc{Direct} & \textbf{60} & 0 & \textbf{0} & 1.391 & 7.552 & 87.75 \\%
   & \textsc{Compact-Base} \textbf{(ours)} & \textbf{60} & 0 & \textbf{0} & 0.922 & \textbf{3.908} & \textbf{69.91} \\%
   & \textsc{Compact-Objective} \textbf{(ours)} & \textbf{60} & 0 & \textbf{0} & \textbf{0.823} & 11.066 & 71.53 \\%
  \addlinespace[1pt]%
  \textsc{IR-IM-IS} & \textsc{Direct} & \textbf{48} & 0 & \textbf{12} & 6.862 & \textbf{3{,}014.146} & 89.06 \\%
   & \textsc{Compact-Base} \textbf{(ours)} & 47 & 0 & 13 & \textbf{1.820} & 3{,}156.525 & 74.65 \\%
   & \textsc{Compact-Objective} \textbf{(ours)} & 47 & 0 & 13 & 3.212 & 3{,}189.273 & \textbf{73.33} \\%
  \addlinespace%
  \multicolumn{8}{@{}l}{\textit{Stress-high ($n=40$)}} \\%
  \textsc{BG-d2} & \textsc{Direct} & 28 & 0 & 12 & 30.691 & 5{,}239.100 & 152.25 \\%
   & \textsc{Compact-Base} \textbf{(ours)} & \textbf{29} & 0 & \textbf{11} & \textbf{27.242} & \textbf{4{,}321.344} & 78.71 \\%
   & \textsc{Compact-Objective} \textbf{(ours)} & 27 & 0 & 13 & 30.179 & 5{,}011.093 & \textbf{71.73} \\%
  \addlinespace[1pt]%
  \textsc{IR} & \textsc{Direct} & \textbf{40} & 0 & \textbf{0} & 2.727 & 4.061 & 89.22 \\%
   & \textsc{Compact-Base} \textbf{(ours)} & \textbf{40} & 0 & \textbf{0} & 0.785 & \textbf{2.116} & \textbf{68.32} \\%
   & \textsc{Compact-Objective} \textbf{(ours)} & \textbf{40} & 0 & \textbf{0} & \textbf{0.688} & 3.688 & 68.33 \\%
  \addlinespace[1pt]%
  \textsc{IR-IM-IS} & \textsc{Direct} & 20 & 0 & 20 & 6{,}957.436 & 7{,}525.762 & 158.37 \\%
   & \textsc{Compact-Base} \textbf{(ours)} & 20 & 0 & 20 & 5{,}816.335 & 7{,}492.235 & 163.82 \\%
   & \textsc{Compact-Objective} \textbf{(ours)} & \textbf{21} & 0 & \textbf{19} & \textbf{4{,}821.423} & \textbf{7{,}436.858} & \textbf{142.65} \\%
}

\newcommand{\StressQualityRows}{%
  Stress & 47 & $\Delta_I$ & 0 & 47 & 0 \\%
   & 47 & $I_{\max}$ & 4 & 43 & 0 \\%
   & 47 & $I_\Sigma$ & 22 & 25 & 0 \\%
  \addlinespace%
  Stress-high & 21 & $\Delta_I$ & 0 & 21 & 0 \\%
   & 21 & $I_{\max}$ & 0 & 21 & 0 \\%
   & 21 & $I_\Sigma$ & 20 & 1 & 0 \\%
}

\newcommand{\AblationRuntimeRows}{%
  \textsc{BG-d2} & \tablecheck &  &  &  &  & \textbf{126} & 0.706 & 2.967 & 76.27 \\%
   & \tablecheck & \tablecheck &  &  &  & \textbf{126} & 0.584 & 3.539 & 72.58 \\%
   & \tablecheck &  & \tablecheck &  &  & \textbf{126} & 0.655 & \textbf{2.775} & 76.04 \\%
   & \tablecheck & \tablecheck & \tablecheck &  &  & \textbf{126} & \textbf{0.571} & 2.964 & 71.77 \\%
   & \tablecheck & \tablecheck & \tablecheck &  & \tablecheck & \textbf{126} & 0.612 & 4.000 & \textbf{66.54} \\%
  \addlinespace%
  \textsc{IR} & \tablecheck &  &  &  &  & \textbf{126} & 0.481 & \textbf{0.801} & 69.21 \\%
   & \tablecheck &  & \tablecheck &  &  & \textbf{126} & 0.498 & 0.847 & 69.01 \\%
   & \tablecheck &  &  & \tablecheck &  & \textbf{126} & \textbf{0.476} & 0.832 & 69.31 \\%
   & \tablecheck &  & \tablecheck & \tablecheck &  & \textbf{126} & 0.492 & 0.916 & 69.17 \\%
   & \tablecheck &  & \tablecheck & \tablecheck & \tablecheck & \textbf{126} & 0.533 & 0.825 & \textbf{66.08} \\%
  \addlinespace%
  \textsc{IR-IM-IS} & \tablecheck &  &  &  &  & 120 & 0.577 & 723.927 & 69.38 \\%
   & \tablecheck &  & \tablecheck &  &  & \textbf{121} & 0.577 & 666.331 & 69.41 \\%
   & \tablecheck &  &  & \tablecheck &  & \textbf{121} & \textbf{0.571} & \textbf{640.149} & 69.51 \\%
   & \tablecheck &  & \tablecheck & \tablecheck &  & \textbf{121} & 0.585 & 661.130 & 69.47 \\%
   & \tablecheck &  & \tablecheck & \tablecheck & \tablecheck & \textbf{121} & 0.591 & 650.678 & \textbf{67.97} \\%
}

\newcommand{\StressEncodingEffectRows}{%
  \multicolumn{6}{@{}l}{\textit{Stress ($n=60$)}} \\%
  \textsc{BG-d2} & \textsc{Direct} $\rightarrow$ \textsc{Compact-Base} \textbf{(ours)} & 44.60 & 52 & 4.16$\times$ & 60/0/0 \\%
   & \textsc{Direct} $\rightarrow$ \textsc{Compact-Objective} \textbf{(ours)} & 64.90 & 51 & 2.40$\times$ & 60/0/0 \\%
   & \textsc{Compact-Base} \textbf{(ours)} $\rightarrow$ \textsc{Compact-Objective} \textbf{(ours)} & 18.66 & 54 & 0.53$\times$ & 60/0/0 \\%
  \addlinespace[1pt]%
  \textsc{IR} & \textsc{Direct} $\rightarrow$ \textsc{Compact-Base} \textbf{(ours)} & 33.91 & 60 & 1.83$\times$ & 60/0/0 \\%
   & \textsc{Direct} $\rightarrow$ \textsc{Compact-Objective} \textbf{(ours)} & 37.73 & 60 & 1.90$\times$ & 60/0/0 \\%
   & \textsc{Compact-Base} \textbf{(ours)} $\rightarrow$ \textsc{Compact-Objective} \textbf{(ours)} & 5.96 & 60 & 1.13$\times$ & 60/0/0 \\%
  \addlinespace[1pt]%
  \textsc{IR-IM-IS} & \textsc{Direct} $\rightarrow$ \textsc{Compact-Base} \textbf{(ours)} & 34.09 & 47 & 2.62$\times$ & 60/0/0 \\%
   & \textsc{Direct} $\rightarrow$ \textsc{Compact-Objective} \textbf{(ours)} & 37.89 & 47 & 1.44$\times$ & 60/0/0 \\%
   & \textsc{Compact-Base} \textbf{(ours)} $\rightarrow$ \textsc{Compact-Objective} \textbf{(ours)} & 5.95 & 47 & 0.55$\times$ & 60/0/0 \\%
  \addlinespace%
  \multicolumn{6}{@{}l}{\textit{Stress-high ($n=40$)}} \\%
  \textsc{BG-d2} & \textsc{Direct} $\rightarrow$ \textsc{Compact-Base} \textbf{(ours)} & 50.97 & 25 & 3.79$\times$ & 40/0/0 \\%
   & \textsc{Direct} $\rightarrow$ \textsc{Compact-Objective} \textbf{(ours)} & 68.24 & 25 & 1.41$\times$ & 40/0/0 \\%
   & \textsc{Compact-Base} \textbf{(ours)} $\rightarrow$ \textsc{Compact-Objective} \textbf{(ours)} & 17.90 & 27 & 0.46$\times$ & 40/0/0 \\%
  \addlinespace[1pt]%
  \textsc{IR} & \textsc{Direct} $\rightarrow$ \textsc{Compact-Base} \textbf{(ours)} & 43.75 & 40 & 2.12$\times$ & 40/0/0 \\%
   & \textsc{Direct} $\rightarrow$ \textsc{Compact-Objective} \textbf{(ours)} & 46.12 & 40 & 2.40$\times$ & 40/0/0 \\%
   & \textsc{Compact-Base} \textbf{(ours)} $\rightarrow$ \textsc{Compact-Objective} \textbf{(ours)} & 4.42 & 40 & 1.14$\times$ & 40/0/0 \\%
  \addlinespace[1pt]%
  \textsc{IR-IM-IS} & \textsc{Direct} $\rightarrow$ \textsc{Compact-Base} \textbf{(ours)} & 43.89 & 18 & 1.47$\times$ & 40/0/0 \\%
   & \textsc{Direct} $\rightarrow$ \textsc{Compact-Objective} \textbf{(ours)} & 46.25 & 19 & 1.13$\times$ & 40/0/0 \\%
   & \textsc{Compact-Base} \textbf{(ours)} $\rightarrow$ \textsc{Compact-Objective} \textbf{(ours)} & 4.41 & 18 & 0.79$\times$ & 40/0/0 \\%
}

\newcommand{\StressAblationFormulaRows}{%
  Stress & \textsc{BG-d2} & \tablecheck &  &  &  &  & 31{,}339.5 & 80{,}066.5 & 194{,}137 & -- \\%
   &  & \tablecheck & \tablecheck &  &  &  & 27{,}254 & 71{,}536.5 & 164{,}158 & 10.78 \\%
   &  & \tablecheck &  & \tablecheck &  &  & 30{,}668 & 78{,}196.5 & 189{,}866 & 2.18 \\%
   &  & \tablecheck & \tablecheck & \tablecheck &  &  & 26{,}581.5 & 69{,}666.5 & 159{,}887 & 13.37 \\%
   &  & \tablecheck & \tablecheck & \tablecheck &  & \tablecheck & 27{,}556.5 & 65{,}562.5 & 151{,}679 & 5.33 \\%
  \addlinespace[1pt]%
   & \textsc{IR} & \tablecheck &  &  &  &  & 26{,}802 & 70{,}995.5 & 163{,}552 & -- \\%
   &  & \tablecheck &  & \tablecheck &  &  & 26{,}601.5 & 70{,}600.5 & 162{,}762 & 0.65 \\%
   &  & \tablecheck &  &  & \tablecheck &  & 26{,}795 & 70{,}974.5 & 163{,}510 & 0.03 \\%
   &  & \tablecheck &  & \tablecheck & \tablecheck &  & 26{,}594.5 & 70{,}579.5 & 162{,}720 & 0.68 \\%
   &  & \tablecheck &  & \tablecheck & \tablecheck & \tablecheck & 27{,}569.5 & 66{,}475.5 & 154{,}512 & 5.26 \\%
  \addlinespace[1pt]%
   & \textsc{IR-IM-IS} & \tablecheck &  &  &  &  & 26{,}802 & 71{,}119.5 & 163{,}676 & -- \\%
   &  & \tablecheck &  & \tablecheck &  &  & 26{,}601.5 & 70{,}724.5 & 162{,}886 & 0.65 \\%
   &  & \tablecheck &  &  & \tablecheck &  & 26{,}795 & 71{,}098.5 & 163{,}634 & 0.03 \\%
   &  & \tablecheck &  & \tablecheck & \tablecheck &  & 26{,}594.5 & 70{,}703.5 & 162{,}844 & 0.68 \\%
   &  & \tablecheck &  & \tablecheck & \tablecheck & \tablecheck & 27{,}569.5 & 66{,}599.5 & 154{,}636 & 5.25 \\%
  \addlinespace%
  Stress-high & \textsc{BG-d2} & \tablecheck &  &  &  &  & 30{,}027.5 & 73{,}728 & 179{,}370.5 & -- \\%
   &  & \tablecheck & \tablecheck &  &  &  & 25{,}986 & 65{,}373 & 150{,}042.5 & 11.74 \\%
   &  & \tablecheck &  & \tablecheck &  &  & 29{,}329 & 71{,}816 & 175{,}023.5 & 2.43 \\%
   &  & \tablecheck & \tablecheck & \tablecheck &  &  & 25{,}287.5 & 63{,}461 & 145{,}695.5 & 15.03 \\%
   &  & \tablecheck & \tablecheck & \tablecheck &  & \tablecheck & 26{,}015 & 60{,}638 & 140{,}049.5 & 3.37 \\%
  \addlinespace[1pt]%
   & \textsc{IR} & \tablecheck &  &  &  &  & 25{,}537.5 & 64{,}861.5 & 149{,}516 & -- \\%
   &  & \tablecheck &  & \tablecheck &  &  & 25{,}307.5 & 64{,}399 & 148{,}589.5 & 0.87 \\%
   &  & \tablecheck &  &  & \tablecheck &  & 25{,}530.5 & 64{,}839 & 149{,}471 & 0.04 \\%
   &  & \tablecheck &  & \tablecheck & \tablecheck &  & 25{,}300.5 & 64{,}376.5 & 148{,}544.5 & 0.91 \\%
   &  & \tablecheck &  & \tablecheck & \tablecheck & \tablecheck & 26{,}028 & 61{,}553.5 & 142{,}898.5 & 3.00 \\%
  \addlinespace[1pt]%
   & \textsc{IR-IM-IS} & \tablecheck &  &  &  &  & 25{,}537.5 & 64{,}985 & 149{,}639.5 & -- \\%
   &  & \tablecheck &  & \tablecheck &  &  & 25{,}307.5 & 64{,}522.5 & 148{,}713 & 0.87 \\%
   &  & \tablecheck &  &  & \tablecheck &  & 25{,}530.5 & 64{,}962.5 & 149{,}594.5 & 0.04 \\%
   &  & \tablecheck &  & \tablecheck & \tablecheck &  & 25{,}300.5 & 64{,}500 & 148{,}668 & 0.90 \\%
   &  & \tablecheck &  & \tablecheck & \tablecheck & \tablecheck & 26{,}028 & 61{,}677 & 143{,}022 & 2.99 \\%
}

\newcommand{\StressAblationRuntimeRows}{%
  Stress & \textsc{BG-d2} & \tablecheck &  &  &  &  & 55 & 5 & 5.236 & 1{,}480.636 & 81.66 \\%
   &  & \tablecheck & \tablecheck &  &  &  & 54 & 6 & 4.355 & 1{,}706.232 & 75.66 \\%
   &  & \tablecheck &  & \tablecheck &  &  & 56 & 4 & 3.573 & 1{,}252.422 & 79.06 \\%
   &  & \tablecheck & \tablecheck & \tablecheck &  &  & 54 & 6 & 4.235 & 1{,}729.857 & 74.32 \\%
   &  & \tablecheck & \tablecheck & \tablecheck &  & \tablecheck & 54 & 6 & 3.801 & 1{,}779.542 & 71.72 \\%
  \addlinespace[1pt]%
   & \textsc{IR} & \tablecheck &  &  &  &  & 60 & 0 & 1.738 & 7.308 & 75.01 \\%
   &  & \tablecheck &  & \tablecheck &  &  & 60 & 0 & 1.297 & 19.855 & 75.30 \\%
   &  & \tablecheck &  &  & \tablecheck &  & 60 & 0 & 1.698 & 5.093 & 73.50 \\%
   &  & \tablecheck &  & \tablecheck & \tablecheck &  & 60 & 0 & 1.504 & 11.863 & 75.07 \\%
   &  & \tablecheck &  & \tablecheck & \tablecheck & \tablecheck & 60 & 0 & 0.785 & 10.842 & 71.66 \\%
  \addlinespace[1pt]%
   & \textsc{IR-IM-IS} & \tablecheck &  &  &  &  & 47 & 13 & 3.070 & 3{,}173.621 & 75.64 \\%
   &  & \tablecheck &  & \tablecheck &  &  & 47 & 13 & 3.198 & 3{,}206.638 & 75.49 \\%
   &  & \tablecheck &  &  & \tablecheck &  & 47 & 13 & 3.841 & 3{,}209.012 & 75.91 \\%
   &  & \tablecheck &  & \tablecheck & \tablecheck &  & 47 & 13 & 3.756 & 3{,}188.215 & 77.19 \\%
   &  & \tablecheck &  & \tablecheck & \tablecheck & \tablecheck & 47 & 13 & 1.925 & 3{,}172.080 & 74.75 \\%
  \addlinespace%
  Stress-high & \textsc{BG-d2} & \tablecheck &  &  &  &  & 27 & 13 & 53.822 & 5{,}006.063 & 78.72 \\%
   &  & \tablecheck & \tablecheck &  &  &  & 28 & 12 & 46.390 & 4{,}732.007 & 76.38 \\%
   &  & \tablecheck &  & \tablecheck &  &  & 28 & 12 & 34.445 & 4{,}792.971 & 80.49 \\%
   &  & \tablecheck & \tablecheck & \tablecheck &  &  & 29 & 11 & 57.561 & 4{,}575.978 & 72.67 \\%
   &  & \tablecheck & \tablecheck & \tablecheck &  & \tablecheck & 27 & 13 & 30.179 & 5{,}011.089 & 71.73 \\%
  \addlinespace[1pt]%
   & \textsc{IR} & \tablecheck &  &  &  &  & 40 & 0 & 1.503 & 3.853 & 69.63 \\%
   &  & \tablecheck &  & \tablecheck &  &  & 40 & 0 & 0.692 & 1.961 & 69.86 \\%
   &  & \tablecheck &  &  & \tablecheck &  & 40 & 0 & 1.495 & 6.000 & 70.80 \\%
   &  & \tablecheck &  & \tablecheck & \tablecheck &  & 40 & 0 & 0.687 & 5.460 & 70.34 \\%
   &  & \tablecheck &  & \tablecheck & \tablecheck & \tablecheck & 40 & 0 & 0.650 & 3.573 & 68.80 \\%
  \addlinespace[1pt]%
   & \textsc{IR-IM-IS} & \tablecheck &  &  &  &  & 18 & 22 & 7{,}200.014 & 8{,}056.789 & 147.04 \\%
   &  & \tablecheck &  & \tablecheck &  &  & 20 & 20 & 4{,}949.473 & 7{,}376.436 & 155.91 \\%
   &  & \tablecheck &  &  & \tablecheck &  & 19 & 21 & 7{,}200.015 & 7{,}972.921 & 134.94 \\%
   &  & \tablecheck &  & \tablecheck & \tablecheck &  & 20 & 20 & 7{,}057.032 & 7{,}658.003 & 134.03 \\%
   &  & \tablecheck &  & \tablecheck & \tablecheck & \tablecheck & 19 & 21 & 7{,}200.017 & 7{,}796.991 & 135.95 \\%
}

\title[Compact MaxSAT Encodings for Balanced Waiting]{Compact MaxSAT Encodings for Balanced Waiting in B2B Meeting Scheduling}

\author{Tuyen Van Kieu}
\email{tuyenkv@vnu.edu.vn}

\author{Long Duc Nguyen}
\email{24021555@vnu.edu.vn}

\author{Minh Duc Dang Vu}
\email{24020241@vnu.edu.vn}

\author{Khanh Van To}
\authornote{Corresponding Author.}
\email{khanhtv@vnu.edu.vn}
\affiliation{%
  \department{Faculty of Information Technology}
  \institution{VNU University of Engineering and Technology}
  \city{Hanoi}
  \country{Vietnam}}

\renewcommand{\shortauthors}{Kieu et al.}

\begin{abstract}
\textbf{Background:} Business-to-business (B2B) meeting schedules can leave participants with long and uneven waits. Earlier break-based B2B models minimize the number of breaks between meetings, treating short and long breaks alike. Our earlier work instead balances participants' total waiting times by minimizing the difference between the largest and smallest totals. However, minimizing this difference alone can still leave participants with long waits.

\textbf{Objectives:} We study how to reduce maximum and total waiting while preserving optimal idle-range balance, and how to encode these objectives compactly.

\textbf{Methods:} Waiting is measured in empty slots between a participant's first and last meetings. The balance measure compares participants who request at least two meetings. We use weighted maximum satisfiability (MaxSAT) for lexicographic optimization: idle range first, maximum idle second, and total idle last. \textsc{Compact-Base} retains our earlier feasibility encoding; \textsc{Compact-Objective} adds objective-specific representations. We prove preservation of feasible schedules, objective costs, and the stated priorities. We compare both encodings with \textsc{Direct}, an independent implementation based on the 2022 MaxSAT model. Encoding comparisons use three repetitions per encoding--objective--instance combination on 226 variants of 20 base instances. For schedule quality, we compare break minimization, idle-range minimization, and the lexicographic objective using the same encoding and feasible set.

\textbf{Results:} Across all nine objective--dataset comparisons, \textsc{Compact-Objective} had the lowest median clause count and lower median peak memory than \textsc{Direct}. Runtime rankings changed when we compared runs with the same solver build and computing environment. On \MainQualityN{} instances solved optimally under all three quality objectives, returned lexicographic schedules preserved the optimal idle range. Relative to returned range-only schedules, maximum idle decreased on \MainQualityMaximumLower{} instances and total idle on \MainQualityTotalLower{}. On the \PriorityQualityN{} instances also solved optimally under range-then-total optimization, its returned schedules matched the three-tier objective on idle range, maximum idle, and total idle.

\textbf{Conclusions:} Lexicographic refinement preserves optimal idle-range balance and uses maximum and then total idle to select among equally balanced schedules. \textsc{Compact-Base} supplied the larger clause reduction relative to \textsc{Direct}; \textsc{Compact-Objective} reduced clauses further.
\end{abstract}

\maketitle
\settopmatter{printacmref=true}

\section{Introduction}
\label{sec:introduction}

The business-to-business (B2B) meeting scheduling problem assigns requested meetings to time slots and locations under availability, capacity, fixed-meeting, and precedence constraints. A feasible schedule can still leave participants with long and uneven waits. We therefore consider both waiting duration and its distribution across participants.

Earlier B2B formulations minimize the total number of \emph{breaks}: contiguous blocks of empty slots between meetings \parencite{bofill2015,bofill2019,bofill2022}. Some also constrain differences in break counts between participants. A~break count distinguishes an uninterrupted agenda from a fragmented one, but assigns the same cost to a one-slot break and a five-slot break. It therefore describes how often waiting occurs without measuring its duration.

We measure a participant's waiting by the number of empty slots between their first and last meetings. We measure \emph{balance} by the difference between the largest and smallest idle-slot totals among participants with at least two meetings. This idle range can be small when waits are uniformly short or uniformly long, a distinction also recognized in studies of equity objectives \parencite{karsu2015,matl2018}.

Our earlier work introduced idle-range minimization and a compact MaxSAT encoding \parencite{nguyen2026compact}. Schedules with the same minimum idle range can have different maximum and total idle. This study examines how to distinguish these schedules while preserving the minimum range, and how to represent the resulting priorities compactly.

We first minimize idle range. Among schedules with the minimum range, we minimize the largest participant idle total; among schedules tied on both criteria, we minimize total idle. This lexicographic objective expresses fixed priorities \parencite{marquessilva2011}, keeping balance first and reducing maximum idle before total idle. We compare it with break and range minimization, and with range followed directly by total idle.

Weighted partial maximum satisfiability (MaxSAT) represents feasibility through hard clauses and costs through weighted soft clauses \parencite{bacchus2021}. \textsc{Compact-Base} retains the feasibility encoding from our earlier work. \textsc{Compact-Objective} extends it with objective-specific reductions and an adaptive encoding of participant non-overlap constraints. Their comparison measures the combined effect of these changes; component ablations assess their individual contributions. Both are compared with \textsc{Direct}, our independent implementation based on the 2022 MaxSAT model \parencite{bofill2022}.

This article makes three contributions. First, it examines refinements of idle-range balance that use waiting duration to distinguish equally balanced schedules. Second, it extends the earlier compact encoding to further waiting objectives and proves that their encoded costs and priority orders match the schedule criteria. These representations let the same feasibility encoding support different waiting preferences. Third, it evaluates encoding efficiency and schedule quality in separate comparisons. Encoding comparisons hold the objective and feasible set fixed, whereas schedule-quality comparisons hold the encoding and feasible set fixed. The evaluation uses 126 benchmark instances and 100 precedence-stress variants from 20 base instances.

\section{Related Work}
\label{sec:related-work}
\label{sec:sat-maxsat-related-work}

\textcite{pesant2015} studied constraint-programming and mixed-integer-programming formulations for B2B meeting scheduling. MaxSAT formulations represented breaks through soft clauses \parencite{bofill2015}, and later studies examined implied constraints \parencite{bofill2019,bofill2022impact}. \textcite{bofill2022} compared these approaches on an extended problem that includes fixed meetings, forbidden slots, and precedence constraints. Break-based formulations may impose a \emph{homogeneity} constraint bounding differences in participants' break counts \parencite{bofill2022impact}. The later compact formulation introduced idle-range minimization \parencite{nguyen2026compact}.

Related allocation problems show why the choice of balance measure matters \parencite{karsu2015,matl2018}. Curriculum scheduling has compared maximum credit load with deviations from a target load \parencite{chiarandini2012}. These studies motivate distinguishing waiting disparity from waiting duration in B2B scheduling. Table~\ref{tab:related-work} summarizes the objectives and balance constraints.

\begin{table}[tbp]
\caption{Objectives and balance constraints in B2B scheduling and related problems.}
\label{tab:related-work}
\centering
\settowidth{\settingcolumnwidth}{Workload-balanced}
\settowidth{\qualitycolumnwidth}{preference violations}
\settowidth{\studiescolumnwidth}{(Chiarandini et al. 2012)}
\begin{tabularx}{\textwidth}{@{}L{\settingcolumnwidth}L{\qualitycolumnwidth}>{\raggedright\arraybackslash}X L{\studiescolumnwidth}@{}}
\toprule
Setting & Quality measures & Objective and balance constraints & Studies \\
\midrule
Break-based B2B scheduling & Break count per~participant & Minimize total breaks; optionally bound their range & \parencite{pesant2015,bofill2022} \\
\addlinespace
Idle-based B2B scheduling & Internal idle slots per~participant & Minimize the range of participant~idle~totals & \parencite{nguyen2026compact} \\
\addlinespace
Curriculum balancing & Credit load per term; preference violations & Minimize maximum load or deviations; add weighted preference costs & \parencite{chiarandini2012,ceschia2014} \\
\addlinespace
Workload-balanced routing & Route workload & Minimize range or maximum workload; use lexicographic minimax & \parencite{matl2018,lehuede2020} \\
\addlinespace
\textbf{B2B waiting optimization} & Break count; internal idle slots & \textbf{Compare breaks and idle range; refine by maximum, then total idle} & \textbf{This article} \\
\bottomrule
\end{tabularx}
\Description{Five rows compare break-based B2B scheduling, idle-based B2B scheduling, academic curriculum balancing, workload-balanced routing, and this study by quality measures, objectives and balance constraints, and representative studies.}
\end{table}

\begin{samepage}
Workload-balanced routing has used lexicographic minimax to compare route durations from longest to shortest \parencite{lehuede2020}. Fixed priorities among different criteria also appear in machine scheduling \parencite{chen2023lexicographic}. Boolean lexicographic optimization can be solved through MaxSAT, pseudo-Boolean optimization, or integer programming \parencite{marquessilva2011}. Here, the priorities apply to idle range, maximum participant idle, and total idle.\par
\end{samepage}

Applications of weighted partial MaxSAT include staff scheduling and train rescheduling \parencite{demirovic2019,lemos2024train}. \textcite{jabs2024} developed an incremental SAT-based method for bi-objective MaxSAT that can enumerate Pareto-optimal solutions. With fixed priorities and bounded criterion values, lexicographic optimization can be represented by one weighted MaxSAT objective \parencite{bacchus2021,marquessilva2011}. We use this representation for the B2B waiting criteria.

Cardinality encodings trade auxiliary variables against clauses \parencite{sinz2005,abio2013}, and specialized pseudo-Boolean encodings exploit at-most-one structure \parencite{bofill2022pbamo}. Commander encodings can reduce the number of at-most-one clauses by introducing auxiliary variables \parencite{klieber2007}. For participant non-overlap, our adaptive configuration selects between established pairwise and commander encodings. We evaluate this choice within the compact B2B formulations.

A shorter version of this work has been accepted for presentation at RIVF 2026,
but its proceedings record was not available when this manuscript was prepared.
The earlier version combined solution-preserving domain filtering, shared-suffix
precedence encoding, and an idle-range objective with span-based idle thresholds \parencite{nguyen2026compact}.
The journal version retains these constructions and extends the cost-preservation
results to further waiting objectives. It combines established lexicographic
optimization and at-most-one encodings with objective-specific reductions for
the B2B waiting criteria. An optional extension, \emph{attainable-value compression},
represents only the idle totals attainable within each participant's meeting domains.
The experiments add matched encoding comparisons with three repetitions,
returned-schedule quality comparisons, and component ablations.

\section{Problem Definition}
\label{sec:problem}

Break counts measure how often waiting occurs; idle-slot counts measure its duration. We define these measures before introducing objectives for balancing waiting and limiting its duration. Table~\ref{tab:problem-notation} summarizes the scheduling and waiting notation.

\begin{table}[tbp]
\caption{Notation for the B2B instance, feasible schedule, and participant-waiting measures.}
\label{tab:problem-notation}
\centering
\begin{tabularx}{\textwidth}{@{}l>{\raggedright\arraybackslash}X@{}}
\toprule
Symbol & Meaning \\
\midrule
\multicolumn{2}{@{}l}{\textit{Instance parameters}} \\
$M,P,T,h$ & Sets of meetings and participants, ordered slot set $T=\{1,\ldots,h\}$, and number of slots $h$ \\
$K$ & Number of interchangeable locations available in each slot \\
$\pi(m),M_p,r_p$ & Two participants in meeting $m$; meetings of participant $p$; and $r_p=|M_p|$ \\
$D_m^0,E,S$ & Initial domain of meeting $m$; direct precedence arc set; and schedule $S:M\rightarrow T$ \\
\addlinespace
\multicolumn{2}{@{}l}{\textit{Participant sets and waiting metrics}} \\
$P^\star$ & Participants requesting at least two meetings: $\{p\in P:r_p\ge2\}$ \\
$G_p(S),I_p(S)$ & Break count and internal idle-slot count of participant $p$ \\
$G_\Sigma(S),I_\Sigma(S),I_{\mathrm{SQ}}(S)$ & Total break count, total internal idle-slot count, and sum of squared individual idle counts \\
$\Delta_G(S),\Delta_I(S)$ & Ranges of break counts over $P$ and internal idle-slot counts over $P^\star$ \\
$I_{\max}(S)$ & Largest participant internal idle-slot count \\
\addlinespace
\multicolumn{2}{@{}l}{\textit{Constraints and feasibility}} \\
$\mathcal F$ & Base feasible set: schedules satisfying constraints (F1)--(F4) \\
$\mathcal F_q$ & Schedules admissible under optimization mode $q$, including any mode-specific hard~condition \\
\bottomrule
\end{tabularx}
\Description{Definitions of the global problem, schedule, and schedule-quality symbols grouped by instance parameters, participant waiting metrics, and feasibility constraints.}
\end{table}

\subsection{Feasible Schedules}

Let $M$, $P$, and $T=\{1,\ldots,h\}$ be the meetings, participants, and ordered slots. Every $m\in M$ involves exactly two distinct participants, written $\pi(m)\subseteq P$, $|\pi(m)|=2$, and occupies one slot. There are $K$ interchangeable locations, and $E\subseteq M\times M$ is a directed set of strict precedence arcs.

A session is a group of consecutive event slots. The initial domain $D_m^0\subseteq T$ contains the slots permitted by the requested session, if specified, and both participants' availability. If meeting $m$ is fixed at slot $t$, its domain is further restricted to $D_m^0\cap\{t\}$.

A slot schedule is a function $S:M\rightarrow T$.  It is feasible exactly when
\begin{align}
 S(m)&\in D_m^0 && \forall m\in M, \label{eq:domain}\tag{F1}\\
 S(m)&\ne S(m') && \forall m,m'\in M:\ m\ne m',\ \pi(m)\cap\pi(m')\ne\emptyset,
   \label{eq:participant-conflict}\tag{F2}\\
 |\{m\in M:S(m)=t\}|&\le K && \forall t\in T, \label{eq:capacity}\tag{F3}\\
 S(i)&<S(j) && \forall(i,j)\in E. \label{eq:strict-precedence}\tag{F4}
\end{align}
Constraint (F1) assigns each meeting to an allowed slot. Constraint (F2) prevents simultaneous meetings for a participant; (F3) limits parallel meetings, and (F4) enforces precedence. Because locations are interchangeable, meetings in the same slot can be assigned to distinct locations whenever their number does not exceed $K$.

\subsection{Breaks and Idle Slots}

We measure waiting only between a participant's first and last meetings. The following definition separates how often waiting occurs from how long it lasts.

\begin{definition}[Breaks and internal idle slots]
\label{def:waiting-profile}
Let $M_p=\{m\in M:p\in\pi(m)\}$, $r_p=|M_p|$, and let the occupied slots of participant $p$ be $t_1<\cdots<t_{r_p}$. Participant non-overlap makes these slots distinct. If $r_p\le1$, set the break count $G_p(S)$ and the internal idle-slot count $I_p(S)$ to zero. If $r_p\ge2$, define
\begin{align}
 I_p(S)&=t_{r_p}-t_1+1-r_p, \label{eq:idle-participant}\\
 G_p(S)&=|\{i\in\{1,\ldots,r_p-1\}:t_{i+1}>t_i+1\}|.
 \label{eq:break-participant}
\end{align}
\end{definition}

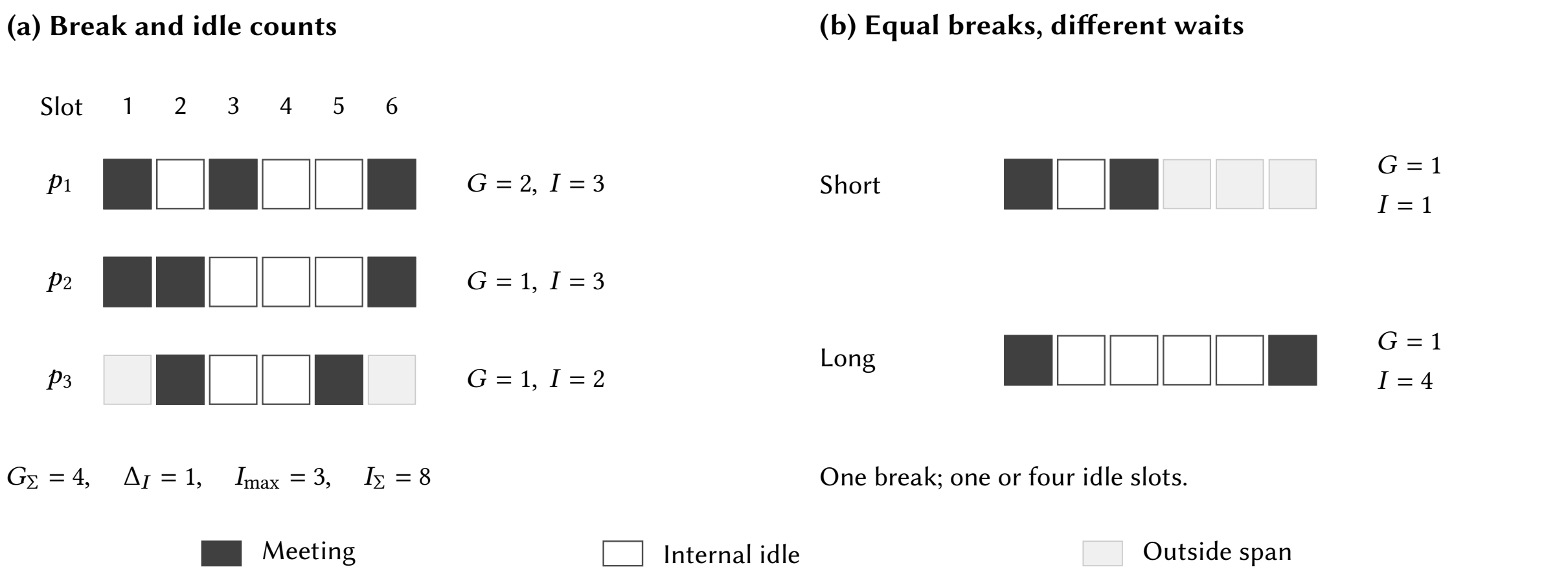
\begin{figure}[htbp]
\centering
\begin{tikzpicture}[x=1mm,y=1mm,every node/.style={fig text}]
\path[use as bounding box] (0,0) rectangle (159,60);
\node[fig heading] (panel-a) at (0,56) {(a) Break and idle counts};
\node[fig heading] (panel-b) at (83,56) {(b) Equal breaks, different waits};
\node[anchor=east] at (8,48) {Slot};
\foreach \t in {1,...,6}{
  \pgfmathsetmacro{\xx}{10+(\t-1)*5.4+2.4}
  \node at (\xx,48) {\t};
}
\foreach \row/\yy/\first/\last/\meetings/\gg/\ii in
 {1/40/1/6/{1,3,6}/2/3,2/30/1/6/{1,2,6}/1/3,3/20/2/5/{2,5}/1/2}{
  \node[anchor=east] at (7,\yy) {$p_{\row}$};
  \foreach \t in {1,...,6}{
    \pgfmathsetmacro{\xx}{10+(\t-1)*5.4}
    \ifnum\t<\first
      \path[fill=black!6,draw=black!20,line width=.35pt] (\xx,\yy-2.5) rectangle +(4.8,5);
    \else\ifnum\t>\last
      \path[fill=black!6,draw=black!20,line width=.35pt] (\xx,\yy-2.5) rectangle +(4.8,5);
    \else
      \path[fill=white,draw=black!70,line width=.45pt] (\xx,\yy-2.5) rectangle +(4.8,5);
    \fi\fi
    \foreach \m in \meetings{\ifnum\t=\m
      \path[fill=black!75,draw=black!75,line width=.45pt] (\xx,\yy-2.5) rectangle +(4.8,5);
    \fi}
  }
  \node[anchor=west] at (47,\yy) {$G=\gg,\ I=\ii$};
}
\node[anchor=west] at (0,10) {$G_\Sigma=4,\quad\Delta_I=1,\quad I_{\max}=3,\quad I_\Sigma=8$};

\foreach \yy/\last/\ii/\name in {40/3/1/Short,22/6/4/Long}{
  \node[anchor=west] at (83,\yy) {\name};
  \foreach \t in {1,...,6}{
    \pgfmathsetmacro{\xx}{102+(\t-1)*5.4}
    \ifnum\t>\last
      \path[fill=black!6,draw=black!20,line width=.35pt] (\xx,\yy-2.5) rectangle +(4.8,5);
    \else
      \path[fill=white,draw=black!70,line width=.45pt] (\xx,\yy-2.5) rectangle +(4.8,5);
    \fi
    \ifnum\t=1
      \path[fill=black!75,draw=black!75,line width=.45pt] (\xx,\yy-2.5) rectangle +(4.8,5);
    \fi
    \ifnum\t=\last
      \path[fill=black!75,draw=black!75,line width=.45pt] (\xx,\yy-2.5) rectangle +(4.8,5);
    \fi
  }
  \node[anchor=west,align=left] at (140,\yy) {$G=1$\\$I=\ii$};
}
\node[anchor=west] at (83,10) {One break; one or four idle slots.};
\path[fill=black!75,draw=black!75,line width=.4pt] (20,1) rectangle +(4,2.5);
\node[anchor=west] at (26,2.25) {Meeting};
\path[fill=white,draw=black!70,line width=.45pt] (61,1) rectangle +(4,2.5);
\node[anchor=west] at (67,2.25) {Internal idle};
\path[fill=black!6,draw=black!20,line width=.35pt] (110,1) rectangle +(4,2.5);
\node[anchor=west] at (116,2.25) {Outside span};
\coordinate (a-sw) at (0,7);\coordinate (a-ne) at (76,51);
\coordinate (b-sw) at (83,7);\coordinate (b-ne) at (159,51);
\ReportFigurePanel{waiting_measures}{a}{a-sw}{a-ne}{panel-a}
\ReportFigurePanel{waiting_measures}{b}{b-sw}{b-ne}{panel-b}
\end{tikzpicture}%
\caption{Schematic participant agendas illustrating breaks and idle slots. Filled cells are meetings; outlined empty cells within the meeting span are internal idle slots. Pale cells lie outside the span. The totals below panel (a) summarize its three agendas; panel (b) compares two agendas with one break each.}
\label{fig:waiting-measures}
\Description{Two panels show schematic participant agendas. Panel (a) summarizes three agendas with idle counts of three, three, and two, and break counts of two, one, and one. Panel (b) compares two agendas with one break but one or four idle slots.}
\end{figure}

The idle count measures the number of empty slots between the first and last meetings; the break count measures how many maximal contiguous blocks these slots form \parencite{bofill2015,bofill2019,bofill2022}. For meetings in slots 1, 3, and 6, the empty slots are 2, 4, and 5, giving $I_p=3$ and $G_p=2$. Figure~\ref{fig:waiting-measures} shows schematic participant agendas and compares agendas with equal break counts but different idle durations.

Let $P^\star=\{p\in P:r_p\ge2\}$. Participants outside $P^\star$ have zero internal idle in every schedule and are excluded from the idle-range comparison. Their zero contribution means total idle can still be summed over $P$. Break-count homogeneity is evaluated over all participants, as in the benchmark formulation \parencite{bofill2022}. We summarize idle slots by their total, range, maximum, and sum of squares, and break counts by their total and range:
\begin{align}
 I_\Sigma(S)&=\sum_{p\in P}I_p(S), &
 \Delta_I(S)&=\max_{p\in P^\star}I_p(S)-\min_{p\in P^\star}I_p(S),
 \label{eq:idle-aggregates}\\
 G_\Sigma(S)&=\sum_{p\in P}G_p(S), &
 \Delta_G(S)&=\max_{p\in P}G_p(S)-\min_{p\in P}G_p(S),
 \label{eq:break-aggregates}\\
 I_{\max}(S)&=\max_{p\in P^\star}I_p(S), &
 I_{\mathrm{SQ}}(S)&=\sum_{p\in P}I_p(S)^2.
 \label{eq:additional-idle-aggregates}
\end{align}
We set $\Delta_I=0$ when $|P^\star|\le1$, and $I_{\max}=0$ when $P^\star=\emptyset$. We omit the schedule argument $(S)$ when the context identifies it.

The three schematic agendas in Figure~\ref{fig:waiting-measures}(a) have idle values $(3,3,2)$ and break counts $(2,1,1)$. For these three agendas, the summaries are
\[
 (G_\Sigma,\Delta_G,\Delta_I,I_\Sigma,I_{\max})=(4,1,1,8,3).
\]

\begin{samepage}
\subsection{Optimization Objectives}
\label{sec:objective-family}
\label{sec:additional-idle-modes}

Each optimization mode specifies an objective and, where applicable, an additional hard constraint. Both break-minimization modes, \textsc{BG-d2} and \textsc{BG-d$\infty$}, minimize $G_\Sigma$. \textsc{BG-d2} additionally requires $\Delta_G\le2$, whereas \textsc{BG-d$\infty$} imposes no break-count homogeneity bound. Let $\mathcal F$ denote the base feasible set defined by constraints~\eqref{eq:domain}--\eqref{eq:strict-precedence}. For an optimization mode $q$ considered in this article, the feasible set is
\begin{equation}
 \mathcal F_q=
 \begin{cases}
  \{S\in\mathcal F:\Delta_G(S)\le2\}, & q=\textsc{BG-d2},\\
  \mathcal F, & \text{otherwise}.
 \end{cases}
\label{eq:mode-feasible-sets}
\end{equation}

The abbreviations IR, IM, and IS denote idle range, maximum idle, and total idle, respectively. The \emph{idle-range objective} (\textsc{IR}) minimizes $\Delta_I$. The \emph{lexicographic waiting objective} (\textsc{IR-IM-IS}) minimizes $(\Delta_I,I_{\max},I_\Sigma)$ in that order.
\end{samepage}

For two participants, idle totals $(1,1)$ and $(2,2)$ both have range zero. Their maximum idle values are 1 and 2, and their totals are 2 and 4. Range minimization alone does not distinguish them. The lexicographic refinement first distinguishes tied ranges by maximum idle, then uses total idle when both earlier criteria tie.

Table~\ref{tab:objective-summary} lists four further waiting objectives. We also consider \textsc{IR-IM-ISQ}, which replaces total idle in the last tier with squared idle and minimizes $(\Delta_I,I_{\max},I_{\mathrm{SQ}})$.

\begin{table}[htbp]
\caption{Additional waiting objectives, all optimized over the base feasible set $\mathcal F$. Multiple criteria are minimized lexicographically from left to right.}
\label{tab:objective-summary}
\centering
\begin{tabularx}{\textwidth}{@{}ll>{\raggedright\arraybackslash}X@{}}
\toprule
Objective & Objective vector & Scheduling preference \\
\midrule
Total idle (\textsc{IS}) & $(I_\Sigma)$ & Minimize total idle \\
Squared idle (\textsc{ISQ}) & $(I_{\mathrm{SQ}})$ & Penalize larger participant idle totals quadratically \\
Maximum then total idle (\textsc{IM-IS}) & $(I_{\max},I_\Sigma)$ & Minimize maximum idle first, then total idle \\
Range then total idle (\textsc{IR-IS}) & $(\Delta_I,I_\Sigma)$ & Minimize idle range first, then total idle \\
\bottomrule
\end{tabularx}
\Description{Four rows list the additional idle-slot objectives, their objective vectors, and the corresponding scheduling preferences.}
\end{table}

\section{MaxSAT Formulations}
\label{sec:compact-framework}

\begin{figure}[htbp]
\centering
\begin{tikzpicture}[x=1mm,y=1mm,every node/.style={fig text}]
\path[use as bounding box] (0,6) rectangle (159,48);
\node[fig box,text width=22mm] (instance) at (14,41) {B2B instance};
\node[fig box,text width=22mm] (objective) at (14,13) {Waiting objective};
\node[fig box,text width=29mm,minimum height=11mm] (feasibility) at (54,41) {\textbf{Domain filtering}\\and feasibility clauses};
\node[fig box,text width=29mm,minimum height=11mm] (cost) at (54,13) {\textbf{Objective clauses}\\(hard and soft)};
\node[fig box,text width=27mm,minimum height=12mm] (formula) at (96,27) {\textbf{Weighted}\\MaxSAT formula};
\node[fig box,text width=24mm] (solver) at (140,38) {MaxSAT solver};
\node[fig box,text width=24mm] (result) at (140,14) {Solver result};
\draw[fig arrow] (instance.east) -- (feasibility.west);
\draw[fig arrow] (objective.east) -- (cost.west);
\draw[fig arrow] (feasibility.south) -- (cost.north);
\node[anchor=east] at (51,27) {Meeting variables};
\draw[fig arrow] (feasibility.east) -| (formula.north);
\draw[fig arrow] (cost.east) -| (formula.south);
\draw[fig arrow] (formula.east) -- ++(5,0) |- (solver.west);
\draw[fig arrow] (solver.south) -- (result.north);
\end{tikzpicture}%
\caption{Construction and solution of a weighted MaxSAT formula. The feasibility and objective clauses are combined before optimization. Algorithm~\ref{alg:compile-wcnf} summarizes the compact construction.}
\label{fig:overall-framework}
\Description{An instance produces feasibility clauses and meeting variables; the selected waiting objective produces hard and soft objective clauses. Both clause sets enter a weighted MaxSAT formula, which is passed to a solver.}
\end{figure}
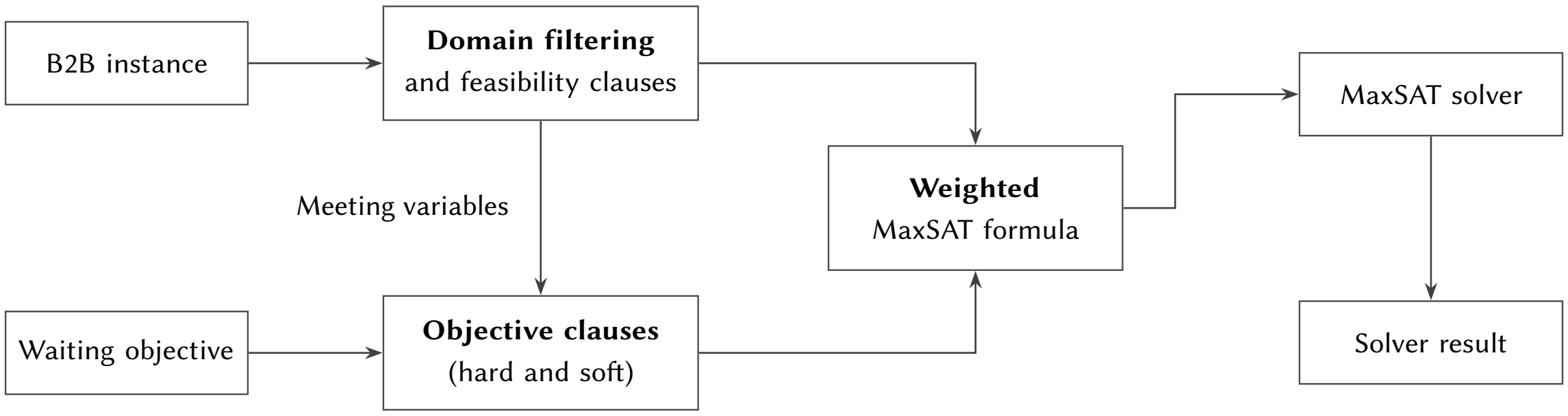

Weighted partial MaxSAT minimizes the cost of violated soft clauses while satisfying every hard clause \parencite{bacchus2021}. We use hard clauses for feasibility and soft clauses for waiting costs. Figure~\ref{fig:overall-framework} shows the construction and solution process.

\subsection{Overview}

We compare three exact MaxSAT encodings. The \emph{direct per-slot encoding} (\textsc{Direct}) is our independent implementation based on the full-domain MaxSAT model of \textcite{bofill2022}, extended here with the idle objectives. For \textsc{BG-d2}, participant non-overlap and break-count thresholds use sequential counters; the published implementation used pairwise clauses and cardinality networks, respectively.

The \emph{base compact encoding} (\textsc{Compact-Base}) retains reduced domains, shared suffix encoding, and span-based idle thresholds from earlier work \parencite{nguyen2026compact}, adapting the objective costs to the criteria studied here. The \emph{objective-aware compact encoding} (\textsc{Compact-Objective}) omits unused objective variables and reuses equivalent states. It also replaces range-indicator penalties with direct soft clauses and applies established pairwise and commander encodings adaptively to participant non-overlap. Table~\ref{tab:formulation-components} summarizes the components. Both compact encodings are marked `ours' in the results tables.

\begin{table}[htbp]
\caption{Structural differences among the three encodings in the matched comparison.}
\label{tab:formulation-components}
\centering
\settowidth{\directcolumnwidth}{General range constraint}
\settowidth{\basecolumnwidth}{Per-slot prefix/suffix chains}
\begin{tabularx}{\textwidth}{@{}lL{\directcolumnwidth}L{\basecolumnwidth}>{\raggedright\arraybackslash}X@{}}
\toprule
Encoding component & \textsc{Direct} & \textsc{Compact-Base} & \textsc{Compact-Objective} \\
\midrule
Meeting domains & All slots; disallowed choices fixed to false & Filtered domains & Same filtered domains \\
Precedence & Repeated clauses for incompatible slot pairs & Shared suffix states for repeated cuts & Same shared suffix states \\
Participant at-most-one & Sequential counters for \textsc{BG-d2}; pairwise clauses for \textsc{IR} and \textsc{IR-IM-IS} & Pairwise clauses & Adaptive pairwise/\allowbreak commander encoding \\
Objective states & Per-slot occupancy and count states & Per-slot prefix/suffix chains & Required states only; equivalent states shared \\
Range costs & Explicit range indicators & Explicit range indicators & Direct soft clauses when the indicator has no~other~use \\
\textsc{BG-d2} homogeneity & General range constraint & General range constraint & Individual bounds if some $r_p\le1$; general~range~otherwise \\
\bottomrule
\end{tabularx}
\Description{Rows compare domains, precedence, participant non-overlap, objective states, range costs and BG-d2 homogeneity. Direct uses sequential counters for BG-d2 and pairwise participant constraints for the two idle objectives. Compact-Objective combines objective reductions with adaptive pairwise and commander constraints.}
\end{table}

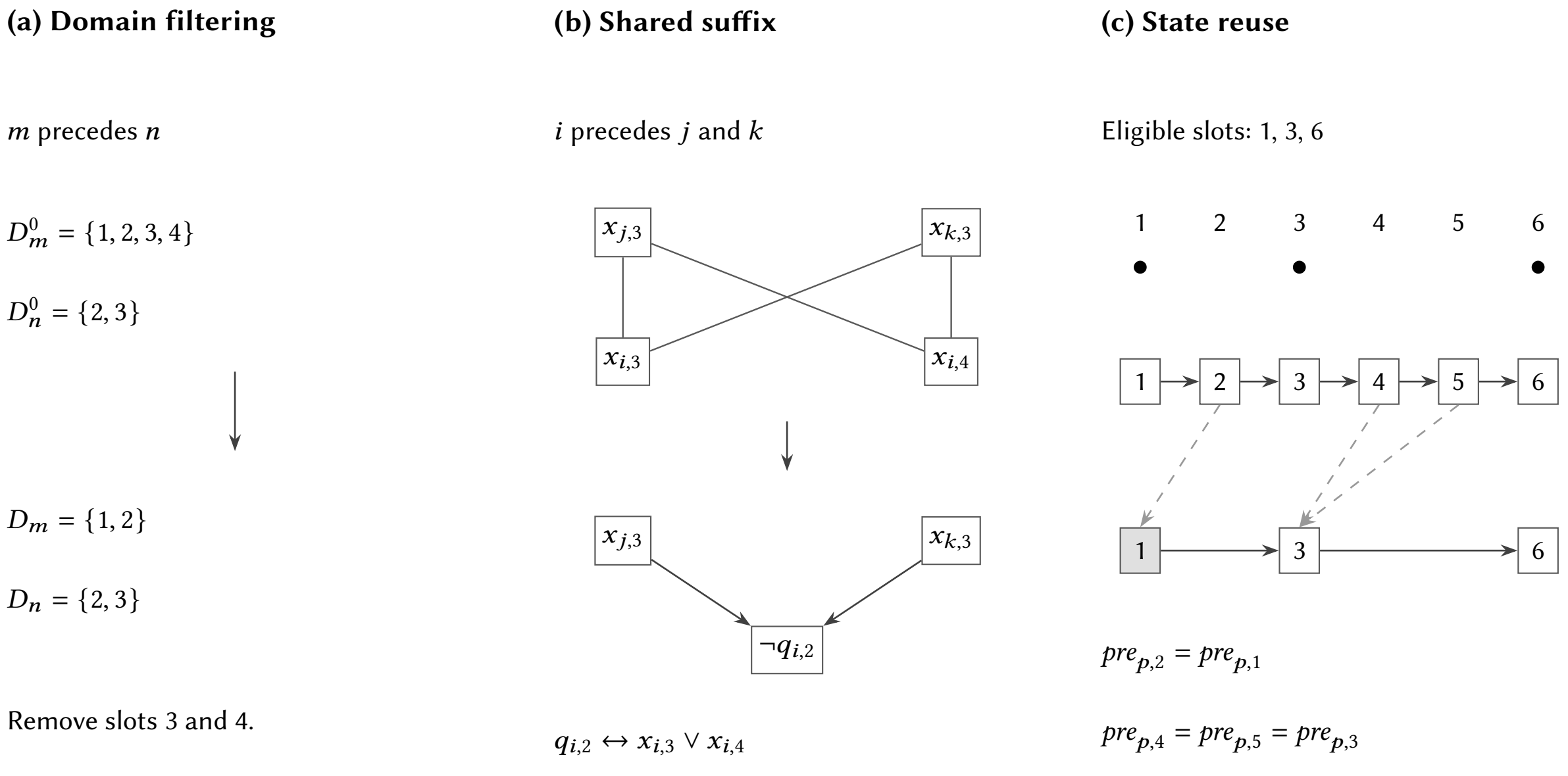
\begin{figure}[tbp]
\centering
\begin{tikzpicture}[x=1mm,y=1mm,every node/.style={fig text}]
\path[use as bounding box] (0,0) rectangle (159,78);
\node[fig heading] (panel-a) at (0,74) {(a) Domain filtering};
\node[fig heading] (panel-b) at (55,74) {(b) Shared suffix};
\node[fig heading] (panel-c) at (110,74) {(c) State reuse};
\node[anchor=west] at (0,63) {$m$ precedes $n$};
\node[anchor=west] at (0,53) {$D_m^0=\{1,2,3,4\}$};
\node[anchor=west] at (0,45) {$D_n^0=\{2,3\}$};
\draw[fig arrow] (23,39) -- (23,31);
\node[anchor=west] at (0,24) {$D_m=\{1,2\}$};
\node[anchor=west] at (0,16) {$D_n=\{2,3\}$};
\node[anchor=west] at (0,4) {Remove slots 3 and 4.};

\node[anchor=west] at (55,63) {$i$ precedes $j$ and $k$};
\node[fig tinybox] (j) at (62,53) {$x_{j,3}$};
\node[fig tinybox] (k) at (95,53) {$x_{k,3}$};
\node[fig tinybox] (i3) at (62,40) {$x_{i,3}$};
\node[fig tinybox] (i4) at (95,40) {$x_{i,4}$};
\draw[fig link] (j) -- (i3);\draw[fig link] (j) -- (i4);
\draw[fig link] (k) -- (i3);\draw[fig link] (k) -- (i4);
\draw[fig arrow] (78.5,34) -- (78.5,29);
\node[fig tinybox] (j2) at (62,22) {$x_{j,3}$};
\node[fig tinybox] (k2) at (95,22) {$x_{k,3}$};
\node[fig tinybox] (suffix) at (78.5,11) {$\neg q_{i,2}$};
\draw[fig arrow] (j2) -- (suffix);\draw[fig arrow] (k2) -- (suffix);
\node[anchor=west] at (55,1.5) {$q_{i,2}\leftrightarrow x_{i,3}\lor x_{i,4}$};

\node[anchor=west] at (110,63) {Eligible slots: 1, 3, 6};
\foreach \t in {1,...,6}{
  \pgfmathsetmacro{\xx}{114+(\t-1)*8}
  \node at (\xx,54) {\t};
  \ifnum\t=1\fill (\xx,49.5) circle (.65mm);\fi
  \ifnum\t=3\fill (\xx,49.5) circle (.65mm);\fi
  \ifnum\t=6\fill (\xx,49.5) circle (.65mm);\fi
  \node[fig tinybox,minimum width=4mm,minimum height=4mm] (q\t) at (\xx,38) {\t};
}
\foreach \t/\u in {1/2,2/3,3/4,4/5,5/6}{\draw[fig arrow] (q\t) -- (q\u);}
\node[fig tinybox,minimum width=4mm,minimum height=4mm,fill=black!12] (r1) at (114,21) {1};
\node[fig tinybox,minimum width=4mm,minimum height=4mm] (r3) at (130,21) {3};
\node[fig tinybox,minimum width=4mm,minimum height=4mm] (r6) at (154,21) {6};
\draw[fig arrow] (r1) -- (r3);\draw[fig arrow] (r3) -- (r6);
\draw[fig arrow,draw=black!40,dashed] (q2.south) -- (r1.north);
\draw[fig arrow,draw=black!40,dashed] (q4.south) -- (r3.north);
\draw[fig arrow,draw=black!40,dashed] (q5.south) -- (r3.north);
\node[anchor=west] at (110,10) {$\mathit{pre}_{p,2}=\mathit{pre}_{p,1}$};
\node[anchor=west] at (110,2) {$\mathit{pre}_{p,4}=\mathit{pre}_{p,5}=\mathit{pre}_{p,3}$};
\coordinate (a-sw) at (0,0);\coordinate (a-ne) at (49,68);
\coordinate (b-sw) at (55,0);\coordinate (b-ne) at (104,68);
\coordinate (c-sw) at (110,0);\coordinate (c-ne) at (159,68);
\ReportFigurePanel{compactness_mechanisms}{a}{a-sw}{a-ne}{panel-a}
\ReportFigurePanel{compactness_mechanisms}{b}{b-sw}{b-ne}{panel-b}
\ReportFigurePanel{compactness_mechanisms}{c}{c-sw}{c-ne}{panel-c}
\end{tikzpicture}%
\caption{Examples of encoding reductions: (a) precedence removes unsupported domain values; (b) with $A_i=(1,2,3,4)$, two successor assignments share the suffix containing slots 3 and 4, starting at zero-based position $c=2$; (c) slots with no eligible meeting reuse prefix states. A prefix state records whether a participant has a meeting at or before that slot.}
\label{fig:compactness-mechanisms}
\Description{Three panels show domain filtering, precedence suffix sharing, and participant-state reuse. In panel b, cut position two in the domain one, two, three, four represents the disjunction of assignments to slots three and four. Panel c reuses prefix states at slots without eligible meetings.}
\end{figure}

Figure~\ref{fig:compactness-mechanisms} illustrates the reductions, and Table~\ref{tab:encoding-notation} defines the encoding symbols. Detailed proofs appear in Appendix~\ref{app:encoding-details}.

\begin{table}[htbp]
\caption{Recurring notation for the feasibility and objective encodings.}
\label{tab:encoding-notation}
\centering
\begin{tabularx}{\textwidth}{@{}l>{\raggedright\arraybackslash}X@{}}
\toprule
Symbol & Meaning \\
\midrule
$D_m,A_m$ & Filtered domain and set of slots with assignment variables for meeting $m$; $A_m=T$ for \textsc{Direct} and $A_m=D_m$ for compact encodings \\
$x_{m,t},y_{p,t}$ & Meeting--slot assignment and participant-occupancy literals \\
$\mathit{pre}_{p,t},\mathit{suf}_{p,t}$ & Indicators of a meeting at or before $t$, and at or after $t$ \\
$f_{p,t},e_{p,t}$ & First-meeting and break-end indicators \\
$\beta_{p,k},\theta_{p,k}$ & Exact break and idle thresholds; compressed idle thresholds use retained levels $v_{p,i}$ \\
$m_k,n_k,\rho_k$ & Indicators that some or all participants reach a count level, and the resulting range indicator; local to each count family \\
$V_p$ & Sorted set of locally attainable idle values for participant $p$ \\
$q_{i,c}$ & Precedence suffix indicator at zero-based position $c$ in the sorted active domain of meeting $i$ \\
$\ell$ & Minimum slot separation on a precedence relation; $\ell=1$ for the strict precedence constraint~(F4) \\
$U_p,U_\Sigma,U_{\max},U_{\mathrm{SQ}}$ & Bounds on individual, total, maximum and squared idle, defined in Section~\ref{sec:idle-slot-encodings} \\
$W_{\mathrm{IR}}^{(3)},W_{\mathrm{IM}}$ & Dominating weights for the range and maximum tiers \\
\bottomrule
\end{tabularx}
\Description{Definitions of the auxiliary symbols used by the proposed Boolean encoding and weighted objectives.}
\end{table}

\subsection{Feasibility Encoding}
\label{sec:compact-components}
\label{sec:filtering}
\label{sec:sparse-suffix}

Domain filtering removes impossible meeting--slot assignments before variables are generated. For each meeting $m$, $A_m$ is the set of slots for which we create assignment literals $x_{m,t}$. \textsc{Direct} uses all slots and fixes disallowed choices to false with unit clauses; the compact encodings use the filtered domain $D_m$.

Filtering first removes slot choices that have no compatible choice for a meeting linked by precedence. For each participant, a bipartite graph connects their meetings to the slots allowed by the current domains. Filtering retains an edge only if it belongs to a matching that assigns all of that participant's meetings to distinct slots \parencite{regin1994,nguyen2026compact}. Finally, if $K$ meetings have a slot as their only remaining choice, capacity excludes that slot for every other meeting.

\begin{theorem}[Domain preservation]
\label{thm:domain-preservation}
Domain filtering terminates. If it reports infeasibility, no feasible schedule exists; otherwise, the returned domain $D_m$ retains $S(m)$ for every meeting $m$ in every feasible schedule $S$.
\end{theorem}

Algorithm~\ref{alg:domain-filtering} and its preservation proof appear in Appendix~\ref{app:filtering-preservation}.

The feasibility clauses encode meeting assignment, participant non-overlap and location capacity:
\begin{align}
 \sum_{t\in A_m}x_{m,t}&=1 &&(m\in M), \label{eq:exactly-one}\\
 \sum_{m\in M_p:t\in A_m}x_{m,t}&\le1 &&(p\in P,t\in T),
 \label{eq:participant-amo}\\
 \sum_{m\in M:t\in A_m}x_{m,t}&\le K &&(t\in T).
 \label{eq:slot-capacity}
\end{align}
Standard cardinality encodings compile these constraints to conjunctive normal form (CNF) \parencite{sinz2005,abio2013,klieber2007}. For each participant--slot at-most-one group, \textsc{Compact-Objective} uses pairwise clauses for fewer than six literals. From six literals onward, it uses a recursive commander encoding with groups of four. The commander encoding groups literals and connects each group to an auxiliary variable. \textsc{Compact-Base} uses pairwise clauses for participant non-overlap throughout. The formulations also retain implied constraints from earlier B2B MaxSAT models \parencite{bofill2019,bofill2022impact}.

Pairwise precedence encoding adds a binary clause for every incompatible pair of slots. Shared suffix encoding represents each forbidden suffix once and reuses it across precedence relations \parencite{nguyen2026compact}. For each pair of meetings connected by a precedence path, the closure records the slot separation implied by the longest such path. Appendix~\ref{app:precedence-preservation} proves the equivalence of these representations.

\begin{proposition}[Precedence equivalence]
\label{prop:precedence-equivalence}
Pairwise precedence clauses and the shared suffix encoding allow exactly the same meeting assignments after auxiliary variables are ignored. Direct precedence arcs and their longest-path closure also admit the same schedules.
\end{proposition}

\subsection{Break Encoding}
\label{sec:objective-encodings}
\label{sec:break-encoding}
\label{sec:additional-idle-encodings}

Break costs require the encoding to identify each maximal block of internal empty slots exactly once. Participant-occupancy literals indicate whether any of the participant's meetings occupies a slot:
\begin{equation}
 y_{p,t}\leftrightarrow
 \bigvee_{m\in M_p:\,t\in A_m}x_{m,t}.
\label{eq:occupancy-channeling}
\end{equation}
The prefix indicator $\mathit{pre}_{p,t}$ is true exactly when the participant has a meeting at or before $t$. The suffix indicator $\mathit{suf}_{p,t}$ is true exactly when the participant has a meeting at or after $t$. Their recurrences start with $\mathit{pre}_{p,0}=0$ and $\mathit{suf}_{p,h+1}=0$. Appendix~\ref{app:objective-identities} gives the hard definitions. At slots without eligible meetings, equivalent neighboring states can share a literal. Such slots still contribute to idle totals when they lie between the first and last meetings.

A break ends at an empty slot $t$ when an earlier meeting exists and slot $t+1$ is occupied:
\begin{equation}
 e_{p,t}\leftrightarrow
 y_{p,t+1}\land\neg y_{p,t}\land\mathit{pre}_{p,t-1},
 \qquad 2\le t\le h-1.
\label{eq:break-end}
\end{equation}
A break-threshold indicator satisfies $\beta_{p,k}\leftrightarrow[G_p\ge k]$. Applying the range construction in Appendix~\ref{app:objective-identities} to these thresholds over $P$ gives the general break-count homogeneity constraint.

\begin{lemma}[Break-end counting]
\label{lem:break-end-counting}
For every participant $p$, the true literals $e_{p,t}$ are in one-to-one correspondence with the breaks in the agenda of $p$. Consequently,
\begin{equation}
 G_p=\sum_{t=2}^{h-1}e_{p,t}
 \quad\text{and}\quad
 G_\Sigma=\sum_{p\in P}\sum_{t=2}^{h-1}e_{p,t}.
\label{eq:break-end-counting}
\end{equation}
\end{lemma}

For \textsc{BG-d2}, the homogeneity constraint simplifies when a participant always has zero breaks, as occurs with at most one meeting. Then $\min_p G_p=0$, so $\Delta_G\le2$ becomes an individual break bound:
\begin{equation}
 \sum_{t=2}^{h-1}e_{p,t}\le2\quad(p\in P).
\label{eq:zero-break-bg}
\end{equation}
We enforce these individual bounds directly and omit the aggregate minimum, maximum and range variables. The objective still minimizes $G_\Sigma$ through unit penalties for break ends.

\begin{proposition}[Zero-break simplification]
\label{prop:zero-break-simplification}
If some participant has zero breaks in every feasible schedule, then $\Delta_G\le2$ holds exactly when $G_p\le2$ for every participant $p$.
\end{proposition}

The implementation applies this simplification when some participant requests at most one meeting, which guarantees zero breaks. Otherwise, \textsc{BG-d2} uses exact break thresholds and the general hard range constraint. \textsc{BG-d$\infty$} needs only unit break-end costs. The ablation study separates the zero-break simplification, omission of unused states, and adaptive participant at-most-one encoding.

\subsection{Idle-Slot Encodings}
\label{sec:idle-slot-encodings}

The total-idle objective can charge each internal empty slot directly. For $p\in P^\star$ and $2\le t\le h-1$, the soft clause
\begin{equation}
 \neg\mathit{pre}_{p,t-1}\ \lor\ y_{p,t}\ \lor\
 \neg\mathit{suf}_{p,t+1}
 \label{eq:direct-idle-clause}
\end{equation}
is violated exactly when slot $t$ is an internal idle slot. Giving every such clause weight one yields $I_\Sigma$ without idle-threshold or first-slot variables.

\begin{lemma}[Direct total-idle cost]
\label{lem:direct-idle-cost}
For a decoded schedule, the number of violated clauses in Equation~\eqref{eq:direct-idle-clause}, summed over participants and slots, equals $I_\Sigma$.
\end{lemma}

The other waiting objectives use exact unary thresholds $\theta_{p,k}\leftrightarrow[I_p\ge k]$. The thresholds are defined from the participant's first and last occupied slots using a first-meeting indicator. Appendix~\ref{app:objective-identities} gives the construction and proves that each threshold has the stated value. The dense representation includes every integer level from $1$ through $U_p$.

For $p\in P^\star$, let $\ell_p$ and $u_p$ be the earliest and latest slots in $\bigcup_{m\in M_p}A_m$. The dense encoding uses the upper bound $U_p=\max\{0,u_p-\ell_p+1-r_p\}$. We use the aggregate bounds
$U_\Sigma=\sum_{p\in P^\star}U_p$, $U_{\max}=\max_{p\in P^\star}U_p$,
and $U_{\mathrm{SQ}}=\sum_{p\in P^\star}U_p^2$. These bounds are computed
separately for each instance, with $U_{\max}=0$ when $P^\star$ is empty.

Exact thresholds also give maximum and range costs. At level $k$, let $m_k$ indicate that some participant in $P^\star$ reaches the level and $n_k$ that all do. The single-literal soft clause $\neg m_k$ contributes to the maximum-idle cost; the soft clause
\begin{equation}
 \neg m_k\lor n_k
 \label{eq:direct-range-cost}
\end{equation}
charges range, because it is violated exactly when some participants reach the level and others do not. These are precisely the integer levels above the minimum idle total and at or below the maximum, so their number equals the range. For example, idle totals 2 and 4 violate the clauses at levels 3 and 4, giving cost 2. The clause replaces a unit penalty on an explicit range indicator. The same construction over $P$ and break thresholds supplies indicators for the hard \textsc{BG-d2} bound. Dense levels use weight one for range and maximum costs.

\begin{lemma}[Threshold counting]
\label{lem:threshold-counting}
If $\theta_{p,k}\leftrightarrow[I_p\ge k]$, then
\begin{equation}
 I_p=\sum_{k=1}^{U_p}\theta_{p,k}.
\label{eq:idle-threshold-counting}
\end{equation}
Using attainable levels with the constants and weights defined in Appendix~\ref{app:objective-identities} preserves $I_p$ and $I_p^2$.
\end{lemma}

The optional attainable-value compression uses the sorted locally possible idle counts $V_p=(v_{p,0},\ldots,v_{p,a_p})$. To find these counts, we consider only that participant's meetings, assigning them to distinct slots in their filtered domains. The compressed representation creates thresholds only for $v_{p,1},\ldots,v_{p,a_p}$; the minimum value $v_{p,0}$ contributes a constant cost. Under compression, $U_p=\max V_p$, and the aggregate bounds use the same formulas above.

Consecutive levels with the same threshold conditions share a range or maximum clause, weighted by the number of levels it represents. Total-idle weights use differences between consecutive retained values; squared-idle weights use differences between their squares. Both costs include the constant contribution from the minimum attainable value. Appendix~\ref{app:objective-identities} defines these costs exactly. The objective-encoding ablation compares dense thresholds with direct idle-slot clauses and attainable-value compression.

\subsection{Objective Encoding and Correctness}
\label{sec:optimization}

We encode the objectives defined in Section~\ref{sec:objective-family} through weighted soft-clause costs. For \textsc{IM-IS}, one weighted MaxSAT call minimizes
\begin{equation}
 (U_\Sigma+1)I_{\max}+I_\Sigma
 \quad\text{over }\mathcal F.
\label{eq:uncapped-idle-weighted-cost}
\end{equation}
For \textsc{IR-IS}, the corresponding cost is $(U_\Sigma+1)\Delta_I+I_\Sigma$. In both cases, a one-unit improvement in the first tier outweighs every possible change in total idle.

For the three-tier objectives, let $U_L$ bound the last tier, where $U_L=U_\Sigma$ for \textsc{IR-IM-IS} and $U_L=U_{\mathrm{SQ}}$ for \textsc{IR-IM-ISQ}. The weights $W_{\mathrm{IM}}=U_L+1$ and $W_{\mathrm{IR}}^{(3)}=(U_{\max}+1)(U_L+1)$ yield the cost
\begin{equation}
 W_{\mathrm{IR}}^{(3)}\Delta_I+W_{\mathrm{IM}}I_{\max}+I_L.
\label{eq:three-tier-weighted-cost}
\end{equation}
Here $I_L$ denotes the selected total-idle or squared-idle final tier. A one-unit improvement in an earlier criterion outweighs every possible change in the later criteria.

\begin{proposition}[Lexicographic weight exactness]
\label{prop:lexicographic-weights}
The scalar costs in Equations~\eqref{eq:uncapped-idle-weighted-cost} and~\eqref{eq:three-tier-weighted-cost} order any two feasible schedules exactly as their stated lexicographic objective vectors.
\end{proposition}

The result follows because each higher-tier weight exceeds the maximum combined contribution of all lower tiers. Appendix~\ref{app:objective-identities} gives the bound argument.

The dense squared-idle cost uses the unary identity
\begin{equation}
 I_{\mathrm{SQ}}
 =\sum_{p\in P^\star}\sum_{k=1}^{U_p}(2k-1)\theta_{p,k}.
\label{eq:squared-idle-unary}
\end{equation}
Under compression, squared-idle costs use differences of squares of consecutive attainable values as weights and include the constant base term. Appendix~\ref{app:objective-identities} gives the resulting identity.

Algorithm~\ref{alg:compile-wcnf} combines the compact feasibility encoding with the selected waiting objective. For \textsc{IS}, the idle representation is either dense thresholds or direct idle-slot clauses; other idle objectives use dense or attainable-value thresholds. The matched encoding comparison uses dense thresholds for \textsc{IR} and \textsc{IR-IM-IS}; the objective-encoding ablation evaluates the alternatives.

\begin{algorithm}[htbp]
\caption{Construct a compact MaxSAT encoding for a waiting objective}
\label{alg:compile-wcnf}
\begin{algorithmic}[1]
\Require B2B instance, objective $q$ (Section~\ref{sec:objective-family}), and idle representation (Section~\ref{sec:idle-slot-encodings})
\Ensure weighted CNF $(\mathcal{H},\mathcal{S},\mathbf{w})$ for $q$, or a report of infeasibility
\State initialize $\mathcal{H}\gets\emptyset$, $\mathcal{S}\gets\emptyset$, and $\mathbf{w}\gets\langle\rangle$
\State filter the meeting domains using Algorithm~\ref{alg:domain-filtering}
\State \Return infeasible if filtering reports infeasibility
\State add assignment, capacity, shared precedence and adaptive non-overlap clauses to $\mathcal{H}$
\State add exact occupancy and the required prefix/suffix definitions to $\mathcal{H}$
\If{$q\in\{\textsc{BG-d2},\textsc{BG-d$\infty$}\}$}
  \State add break definitions and, for \textsc{BG-d2}, homogeneity clauses to $\mathcal{H}$ (Section~\ref{sec:objective-encodings})
  \State form the $G_\Sigma$ soft cost as in Section~\ref{sec:objective-encodings}
\Else
  \State compute bounds and encode the selected idle representation (Section~\ref{sec:idle-slot-encodings})
  \State form the component soft costs for $q$, including base constants (Appendix~\ref{app:objective-identities})
\EndIf
\State scale component soft weights by their priority weights (Section~\ref{sec:optimization}; one for a single criterion) and collect the clauses and weights in $\mathcal{S}$ and $\mathbf{w}$
\State \Return $(\mathcal{H},\mathcal{S},\mathbf{w})$
\end{algorithmic}
\end{algorithm}

\begin{samepage}
Projection onto the meeting variables means discarding auxiliary variables and reading the unique selected slot of each meeting. The following theorem establishes that this decoding preserves feasibility and objective values.

\begin{theorem}[Schedule--encoding correspondence]
\label{prop:objective-preservation}
For every objective $q$ defined in Section~\ref{sec:objective-family}, projecting an assignment satisfying all hard clauses onto the meeting variables yields a schedule in $\mathcal F_q$. Its encoded objective vector equals the vector computed from that schedule. Conversely, every schedule in $\mathcal F_q$ extends to an assignment satisfying all hard clauses with the same vector. Therefore, an optimal weighted-CNF model decodes to an optimal schedule for mode $q$.
\end{theorem}

Appendix~\ref{app:framework-correctness} combines feasibility preservation and the counting identities to prove this correspondence.
\end{samepage}

\Needspace{14\baselineskip}
\section{Experimental Design}
\label{sec:experimental-methodology}

The experiments separate encoding efficiency, schedule quality, and the contribution of individual encoding components.

\subsection{Research Questions}

\emph{RQ1.} How do the three encodings compare in formula size, runtime, and memory under the same objective across the benchmark and precedence-stress sets?

\emph{RQ2.} How do waiting objectives trade off balance, maximum idle, and total idle?

\emph{RQ3.} How do the encoding components affect clause counts, literal occurrences, and variable counts?

\subsection{Benchmark Instances}

All-126 contains 126 B2B benchmark instances derived from 20 base instances \parencite{bofill2022}. These include availability, fixed-slot, and precedence variants, with 125--302 meetings, 42--78 participants, and 8--22 slots.

Stress (60 instances) and Stress-high (40 instances) add precedence constraints to the same 20 base instances. For each participant $p$, the parameter $\gamma$ determines the number of added direct arcs, $\lfloor\gamma |M_p|/100\rfloor$. Each arc follows the order of meetings in a fixed feasible schedule. Stress uses $\gamma\in\{30,35,40\}$ and Stress-high uses $\gamma\in\{50,60\}$. Larger values of $\gamma$ retain every arc from smaller values, so the generating schedule remains feasible throughout. The 226 instances are related variants of 20 base instances; we report descriptive comparisons across these variants.

\subsection{Experimental Setup}

RQ1 compares \textsc{Direct}, \textsc{Compact-Base}, and \textsc{Compact-Objective} under each of \textsc{BG-d2}, \textsc{IR}, and \textsc{IR-IM-IS}. Each comparison fixes the objective and its feasible set: $\mathcal F_{\textsc{BG-d2}}$ for \textsc{BG-d2}, and $\mathcal F$ for the two idle objectives. The idle objectives use dense thresholds in this comparison.

RQ2 compares break, idle-range, and lexicographic waiting objectives on All-126, and the two range-first objectives \textsc{IR} and \textsc{IR-IM-IS} on Stress and Stress-high. Additional comparisons examine total versus squared idle and the order of the waiting priorities. All quality comparisons use \textsc{Compact-Objective} and the same feasible set $\mathcal F$.

\textsc{BG-d$\infty$}, \textsc{IS}, \textsc{ISQ}, \textsc{IM-IS}, and \textsc{IR-IS} each have one run per instance. \textsc{IR}, \textsc{IR-IM-IS}, and \textsc{IR-IM-ISQ} each have three repetitions. Each quality comparison uses instances solved to optimality under every objective being compared.

RQ3 compares component configurations under a fixed objective, using one run per instance and configuration. The objective-encoding ablation compares dense thresholds with direct idle-slot clauses and attainable-value compression.

We implemented the encodings using PySAT \parencite{ignatiev2018}. We used UWrMaxSAT \parencite{piotrow2020} with a two-hour time limit per run. The encoding comparison used Google Cloud virtual machines with 8 vCPUs and 16 GB of memory. Each encoding--objective--instance combination was run three times, giving 6{,}102 runs. Solver builds and software environments varied across runs; we also examined runtimes using runs with matching execution settings.

\subsection{Evaluation Metrics and Analysis}
\label{sec:evaluation-metrics}

Formula size is measured by variables, clauses, and literal occurrences; clause counts include hard and soft clauses. Percentage reductions are computed per paired instance, then summarized by their median. Outcome counts refer to instances. OPT denotes an optimum proved in at least one repetition; UNSAT denotes proved infeasibility. TO means that all repetitions timed out, and Solved equals OPT plus UNSAT.

Runtime includes input parsing, encoding, optimization, and schedule decoding and checking. Peak resident set size (RSS) measures the combined resident memory of the run's processes, in MiB. For each instance, we take the median over repetitions that proved optimality; if none did, we use all three repetitions. Tables report the median of these per-instance values across the data set.

Penalized average runtime (PAR-2) assigns twice the time limit, 14{,}400 seconds, to each timeout. We take the median score over all three repetitions per instance, then average across instances. Paired runtime ratios use median times on instances where both encodings proved optimality in all three repetitions.

To examine the effect of execution settings, we repeat the paired-runtime comparison using runs with matching solver and environment records for the same instance and repetition. Pairs with incomplete records are excluded. Within each environment and across all retained pairs, we take median runtimes for each encoding and instance, then the median \textsc{Compact-Base}/\textsc{Compact-Objective} ratio across instances. The repository lists the matching criteria and sample sizes.

Schedule quality is measured by $G_\Sigma$, $\Delta_I$, $I_{\max}$, $I_\Sigma$, and $I_{\mathrm{SQ}}$, with $\Delta_I$ as the primary balance measure. We also report the population standard deviation (SD) of participant idle totals,
\begin{equation}
 \sigma_I=\sqrt{\frac{I_{\mathrm{SQ}}}{n}-\left(\frac{I_\Sigma}{n}\right)^2},
 \qquad n=|P^\star|.
\end{equation}
Here $P^\star$ contains participants with at least two meetings. We set $\sigma_I=0$ when $P^\star$ is empty. SD describes variation in waiting across participants within a selected schedule; \textsc{ISQ} instead minimizes the sum of squared idle totals. We compute each measure, including $\sigma_I$, for each selected schedule and report its median across instances.

For each instance and objective, we select the schedule from the lowest-numbered repetition that proved optimality. A paired difference is the measure under the first named objective minus its value under the second, on the same instance. Negative, zero, and positive differences give the lower, equal, and higher instance counts, respectively. Optimal schedules for the same objective can differ in measures that the objective does not optimize. Values tied for the best result at the displayed precision are highlighted together. Numerical comparisons in the text use unrounded values.

On small instances, we checked encoded objective values against values computed by enumerating feasible schedules. We checked returned schedules against constraints~(F1)--(F4), their objective values, and the additional break-balance constraint $\Delta_G\le2$ for \textsc{BG-d2}. We also compared formula counts across repetitions and optimal objective vectors across encodings for the same instance and objective.

\FloatBarrier

\section{Results}
\label{sec:results}

We report encoding efficiency (RQ1), schedule quality (RQ2), and component contributions (RQ3).

\subsection{Encoding Size and Solver Performance (RQ1)}

We first compare encoding sizes, then solver outcomes, runtime, and memory under each fixed objective.

\subsubsection{Formula Size}

Both compact encodings used fewer clauses than \textsc{Direct} on every instance compared. \textsc{Compact-Objective} produced the smallest median under all three objectives and on all three data sets (Table~\ref{tab:main-formula-clauses}).

\begin{table}[htbp]
\centering
\caption{Median total clause counts on the three data sets. Bold numerical values mark the smallest displayed median within each data-set--objective block, including rounding ties.}
\label{tab:main-formula-clauses}
\begin{tabular*}{\textwidth}{@{\extracolsep{\fill}}lrrr@{}}
\toprule
Encoding & \textsc{BG-d2} & \textsc{IR} & \textsc{IR-IM-IS} \\
\midrule
\MatchedFormulaRows
\bottomrule
\end{tabular*}
\Description{Median total clause counts for three encodings and three objectives on the benchmark and two precedence-stress sets.}
\end{table}

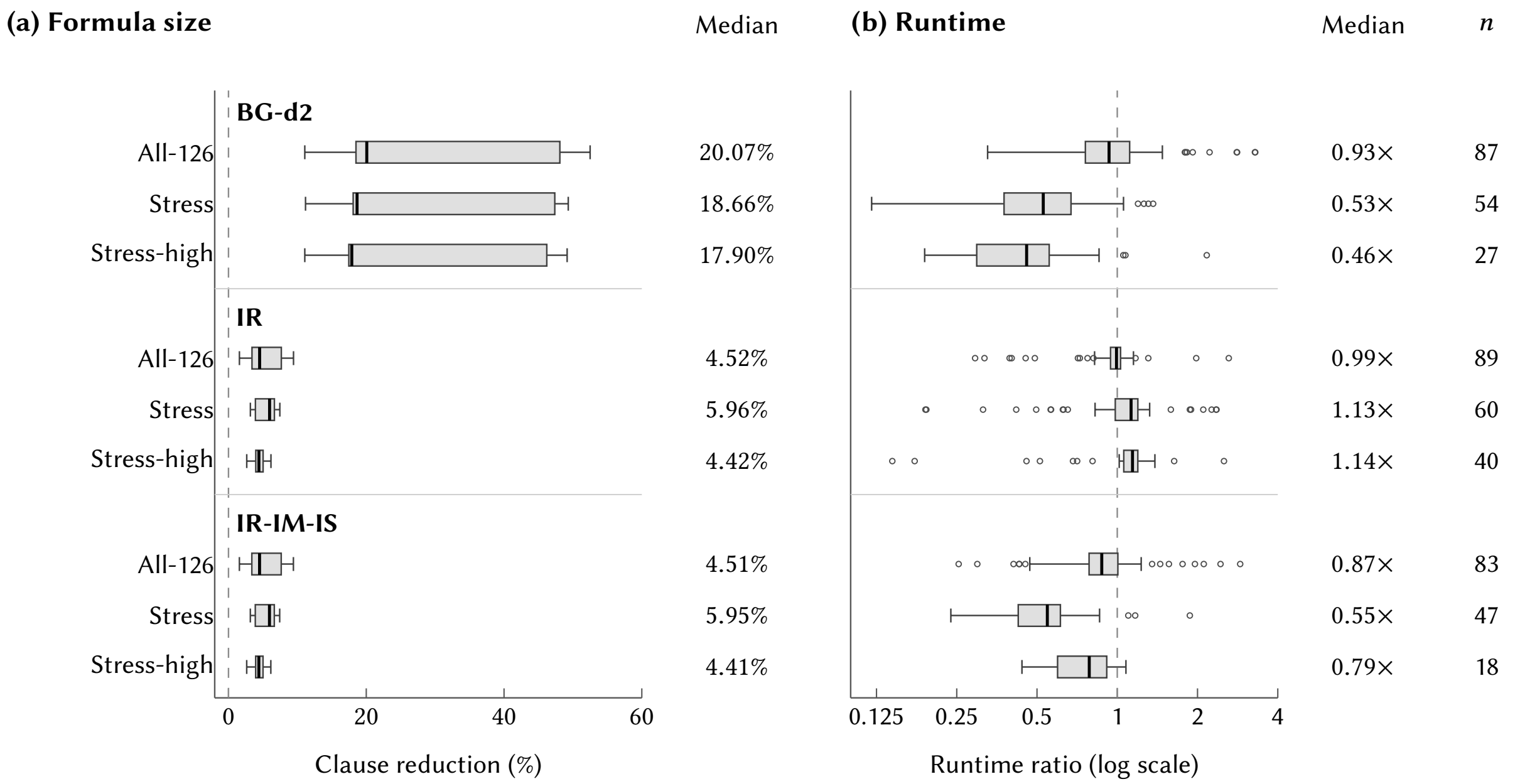
\begin{figure}[!htbp]
\centering
\begingroup%
\expandafter\def\csname EncodingOutliersformula0\endcsname{}
\expandafter\def\csname EncodingOutliersformula1\endcsname{}
\expandafter\def\csname EncodingOutliersformula2\endcsname{}
\expandafter\def\csname EncodingOutliersformula3\endcsname{}
\expandafter\def\csname EncodingOutliersformula4\endcsname{}
\expandafter\def\csname EncodingOutliersformula5\endcsname{}
\expandafter\def\csname EncodingOutliersformula6\endcsname{}
\expandafter\def\csname EncodingOutliersformula7\endcsname{}
\expandafter\def\csname EncodingOutliersformula8\endcsname{}
\expandafter\def\csname EncodingOutliersruntime0\endcsname{(0,1.79342117954453) (1,1.80163694261928) (2,1.82597890605055) (3,1.91905495966976) (4,2.21922773000645) (5,2.80523981221695) (6,2.81557774877649) (7,3.28313602426658) (8,3.29322083617927)}
\expandafter\def\csname EncodingOutliersruntime1\endcsname{(0,1.19603614906589) (1,1.25858326623108) (2,1.30819730122529) (3,1.36335576768737)}
\expandafter\def\csname EncodingOutliersruntime2\endcsname{(0,1.05360989457767) (1,1.07317809659039) (2,2.16593530696374)}
\expandafter\def\csname EncodingOutliersruntime3\endcsname{(0,0.29295193826805) (1,0.31765841930929) (2,0.39474947270608) (3,0.40178531972248) (4,0.453249776022943) (5,0.490778972862513) (6,0.712009560848726) (7,0.724093599910644) (8,0.773585155376434) (9,0.814059363609354) (10,1.17089730035629) (11,1.3073665396846) (12,1.97834502538907) (13,2.61975365867186)}
\expandafter\def\csname EncodingOutliersruntime4\endcsname{(0,0.190755865381634) (1,0.192318701958291) (2,0.313768408423643) (3,0.417881145647434) (4,0.496564482871812) (5,0.562485157038532) (6,0.564486058807214) (7,0.625048070687234) (8,0.628479539355609) (9,0.652264977164154) (10,1.58843430466471) (11,1.87254868665573) (12,1.89118063486427) (13,2.10182125040532) (14,2.25916668464118) (15,2.34774437862014) (16,2.35650422378887)}
\expandafter\def\csname EncodingOutliersruntime5\endcsname{(0,0.143317616172342) (1,0.173811184149368) (2,0.456626936328949) (3,0.512792942648435) (4,0.682196913939751) (5,0.707280414220228) (6,0.807918787774011) (7,1.63486221504385) (8,2.51331003520896)}
\expandafter\def\csname EncodingOutliersruntime6\endcsname{(0,0.254183859648698) (1,0.29856973930688) (2,0.40830062281754) (3,0.428798034364204) (4,0.428848453770522) (5,0.452025294509485) (6,1.35096727164794) (7,1.44902808410678) (8,1.56222115019403) (9,1.75792299227341) (10,1.95600499538387) (11,2.11141114850362) (12,2.43956076553619) (13,2.89110981143183)}
\expandafter\def\csname EncodingOutliersruntime7\endcsname{(0,1.10003247613431) (1,1.16758606586349) (2,1.8718206056886)}
\expandafter\def\csname EncodingOutliersruntime8\endcsname{}
\pgfplotstableread[col sep=comma,row sep=\\]{%
objective,dataset,y,n,lower_whisker,q1,median,q3,upper_whisker,outliers\\
BG-d2,All-126,10,126,11.076887976272,18.5063378836259,20.0747326834586,48.1032739578534,52.4966826740806,0\\
BG-d2,Stress,9,60,11.1828446289923,18.0978730440835,18.6637183557411,47.3673835834784,49.3223745054273,0\\
BG-d2,Stress-high,8,40,11.0772357723577,17.466608893941,17.8971497404649,46.2061748400751,49.1592807523201,0\\
IR,All-126,6,126,1.58599227625511,3.41099559499385,4.52004270970033,7.66935964261068,9.43906643681209,0\\
IR,Stress,5,60,3.17759853903515,3.90829161175212,5.96386712765483,6.66606493733113,7.45436781098725,0\\
IR,Stress-high,4,40,2.64975428343738,3.9609846636031,4.41536423896875,5.02473581867673,6.16605375196067,0\\
IR-IM-IS,All-126,2,126,1.58189977561705,3.38855098216384,4.5095909178418,7.65389415330116,9.42319657915988,0\\
IR-IM-IS,Stress,1,60,3.17093913254769,3.88996701999519,5.94772815380035,6.65184594445462,7.43152428630612,0\\
IR-IM-IS,Stress-high,0,40,2.64422280393651,3.94409610841894,4.40702872581546,5.01524901335574,6.14689206146892,0\\
}\FormulaStats
\pgfplotstableread[col sep=comma,row sep=\\]{%
objective,dataset,y,n,lower_whisker,q1,median,q3,upper_whisker,outliers\\
BG-d2,All-126,10,87,0.32639009828778,0.759280336451219,0.931530895073633,1.11290891469045,1.47656503376873,9\\
BG-d2,Stress,9,54,0.119871718583573,0.376217838214521,0.527774674722277,0.670749584380041,1.05542852239812,4\\
BG-d2,Stress-high,8,27,0.189514917633182,0.297071383262399,0.45710835078377,0.555561068777655,0.854399879588086,3\\
IR,All-126,6,89,0.822096374540027,0.943518994564264,0.993406176219559,1.02895954243418,1.15017390262101,14\\
IR,Stress,5,60,0.825514539230325,0.982932789369041,1.12675704249392,1.19464502781562,1.32270200266891,17\\
IR,Stress-high,4,40,1.01709938432597,1.05777486895794,1.14063899299495,1.19425234738192,1.38415172324715,9\\
IR-IM-IS,All-126,2,83,0.469773184967607,0.784001539352031,0.87470792831289,1.00521484394562,1.22954135530653,14\\
IR-IM-IS,Stress,1,47,0.237596142928574,0.425165503332303,0.546641281968596,0.611972968003661,0.858255582970282,3\\
IR-IM-IS,Stress-high,0,18,0.43908437133348,0.597212222513035,0.785266031225582,0.913236658024869,1.07777819008463,0\\
}\RuntimeStats
\def\DrawEffectBoxes#1#2{%
  \foreach \rr in {0,...,8}{%
    \pgfplotstablegetelem{\rr}{y}\of#2\let\BoxY\pgfplotsretval
    \pgfplotstablegetelem{\rr}{lower_whisker}\of#2\let\BoxLo\pgfplotsretval
    \pgfplotstablegetelem{\rr}{q1}\of#2\let\BoxQOne\pgfplotsretval
    \pgfplotstablegetelem{\rr}{median}\of#2\let\BoxMed\pgfplotsretval
    \pgfplotstablegetelem{\rr}{q3}\of#2\let\BoxQThree\pgfplotsretval
    \pgfplotstablegetelem{\rr}{upper_whisker}\of#2\let\BoxHi\pgfplotsretval
    \pgfplotstablegetelem{\rr}{outliers}\of#2\let\BoxOut\pgfplotsretval
    \ifnum\BoxOut>0
      \edef\BoxData{coordinates {\csname EncodingOutliers#1\rr\endcsname};}%
    \else
      \def\BoxData{coordinates {}; }%
    \fi
    \edef\BoxCommand{\noexpand\addplot[
      boxplot prepared={draw position=\BoxY,lower whisker=\BoxLo,
        lower quartile=\BoxQOne,median=\BoxMed,upper quartile=\BoxQThree,
        upper whisker=\BoxHi},
      boxplot/draw direction=x,boxplot/box extend=.42,boxplot/whisker extend=.27,
      solid,draw=black!75,fill=black!12,line width=.45pt,
      boxplot/every median/.style={draw=black,line width=.9pt},
      mark=o,mark size=.8pt,mark options={draw=black!70,fill=white,line width=.35pt}]
      \BoxData
    }%
    \BoxCommand
  }%
}%
\begin{tikzpicture}[x=1mm,y=1mm,every node/.style={fig text}]
\path[use as bounding box] (0,0) rectangle (159,88);
\node[fig heading] (panel-a) at (0,84) {(a) Formula size};
\node[fig heading] (panel-b) at (89,84) {(b) Runtime};
\node at (77,84) {Median};
\node at (143,84) {Median};
\node at (156,84) {$n$};

\begin{axis}[figure axis,name=formula-axis,at={(22mm,13mm)},anchor=south west,
  width=45mm,height=64mm,scale only axis,
  xmin=-2,xmax=60,ymin=-.6,ymax=11.2,
  xtick={0,20,40,60},ytick={10,9,8,6,5,4,2,1,0},
  yticklabels={All-126,Stress,Stress-high,All-126,Stress,Stress-high,All-126,Stress,Stress-high},
  ytick style={draw=none},xlabel={Clause reduction (\%)},
]
\draw[draw=black!45,dashed,line width=.45pt] (axis cs:0,-.6) -- (axis cs:0,11.2);
\DrawEffectBoxes{formula}{\FormulaStats}
\pgfplotsextra{%
\foreach \yy/\objective in {10.8/BG-d2,6.8/IR,2.8/IR-IM-IS}{
  \node[anchor=west,font=\sffamily\bfseries\fontsize{8}{9.5}\selectfont]
    at (axis cs:1,\yy) {\objective};
}
\foreach \yy in {7.35,3.35}{\draw[draw=black!20,line width=.35pt]
  (axis cs:-2,\yy)--(axis cs:60,\yy);}
}
\end{axis}

\begin{axis}[figure axis,name=runtime-axis,at={(89mm,13mm)},anchor=south west,
  width=45mm,height=64mm,scale only axis,
  xmode=log,log basis x=2,xmin=.1,xmax=4,ymin=-.6,ymax=11.2,
  xtick={.125,.25,.5,1,2,4},xticklabels={0.125,0.25,0.5,1,2,4},ytick=\empty,
  xlabel={Runtime ratio (log scale)},
]
\draw[draw=black!45,dashed,line width=.45pt] (axis cs:1,-.6) -- (axis cs:1,11.2);
\DrawEffectBoxes{runtime}{\RuntimeStats}
\pgfplotsextra{%
\foreach \yy in {7.35,3.35}{\draw[draw=black!20,line width=.35pt]
  (axis cs:.1,\yy)--(axis cs:4,\yy);}
}
\end{axis}

\foreach \rr in {0,...,8}{
  \pgfplotstablegetelem{\rr}{y}\of\FormulaStats\let\BoxY\pgfplotsretval
  \pgfmathsetmacro{\TextY}{13+(\BoxY+.6)/11.8*64}
  \pgfplotstablegetelem{\rr}{median}\of\FormulaStats\let\Value\pgfplotsretval
  \node at (77,\TextY) {\pgfmathprintnumber[fixed,fixed zerofill,precision=2]{\Value}\%};
  \pgfplotstablegetelem{\rr}{median}\of\RuntimeStats\let\Value\pgfplotsretval
  \node at (143,\TextY) {\pgfmathprintnumber[fixed,fixed zerofill,precision=2]{\Value}$\times$};
  \pgfplotstablegetelem{\rr}{n}\of\RuntimeStats\let\Count\pgfplotsretval
  \node at (156,\TextY) {\Count};
}
\coordinate (a-sw) at (formula-axis.south west);\coordinate (a-ne) at (formula-axis.north east);
\coordinate (b-sw) at (runtime-axis.south west);\coordinate (b-ne) at (runtime-axis.north east);
\ReportFigurePanel{matched_encoding_effects}{a}{a-sw}{a-ne}{panel-a}
\ReportFigurePanel{matched_encoding_effects}{b}{b-sw}{b-ne}{panel-b}
\end{tikzpicture}%
\endgroup%
\caption{Effects of \textsc{Compact-Objective} relative to \textsc{Compact-Base}: (a) per-instance clause reductions on all instances within each data set and (b) runtime ratio $t_{\mathrm{Base}}/t_{\mathrm{Objective}}$ on instances solved to optimality in all three repetitions by both encodings ($n$ shown). Runtime summaries combine runs from different software environments. Positive reductions and ratios above one favor \textsc{Compact-Objective}; ratios below one favor \textsc{Compact-Base}. Boxes show medians and interquartile ranges (IQR); whiskers reach the most extreme values within 1.5 IQR of the box, and circles mark outliers.}
\label{fig:encoding-effects}
\Description{Two aligned panels compare the base and objective-aware compact encodings under three objectives on All-126, Stress, and Stress-high. Medians and runtime sample sizes appear in separate columns. The runtime panel combines runs from different software environments.}
\end{figure}

The larger clause reduction came from \textsc{Compact-Base}. For \textsc{IR-IM-IS} on All-126, its median per-instance clause reduction relative to \textsc{Direct} was 40.33\%. \textsc{Compact-Objective} reduced clauses by a further median of 4.51\% per instance relative to \textsc{Compact-Base}. Relative to \textsc{Compact-Base}, it also reduced literal occurrences while increasing the median variable count under both idle objectives. The additional clause savings from \textsc{Compact-Objective} were largest under \textsc{BG-d2} on all three data sets (Figure~\ref{fig:encoding-effects}(a)).

\FloatBarrier
\subsubsection{Runtime and Memory}

Table~\ref{tab:main-runtime-summary} compares outcomes, PAR-2, and memory for all three encodings.

\begin{table}[!htbp]
\centering
\caption{Solver outcomes, PAR-2, and peak RSS for the three encodings. Bold numerical values mark the largest OPT count when counts differ and the lowest displayed PAR-2 or RSS, including rounding ties.}
\label{tab:main-runtime-summary}
\begingroup
\setlength{\tabcolsep}{7pt}
\begin{tabular}{@{}llrrr@{}}
\toprule
Objective & Encoding & OPT/UNSAT/TO & PAR-2 (s) & RSS (MiB) \\
\midrule
\MatchedRuntimePrimaryRows
\bottomrule
\end{tabular}
\endgroup
\par\smallskip\noindent\parbox{\textwidth}{\textit{Note:} OPT requires at least one optimal repetition. PAR-2 uses all three repetitions; RSS uses optimal repetitions when available. Runs used different solver builds and software environments.}
\Description{Counts of optimal, infeasible, and timeout instances, mean PAR-2, and median peak resident set size for Direct, Compact-Base, and Compact-Objective, grouped by data set and objective. Direct is an independent implementation based on the 2022 MaxSAT model. The table combines runs from different solver builds and software environments.}
\end{table}

Compared with \textsc{Direct}, \textsc{Compact-Objective} had lower median RSS in all nine data-set--objective pairs and lower PAR-2 in six. PAR-2 was higher for \textsc{IR} on Stress and for \textsc{IR-IM-IS} on All-126 and Stress. Its OPT count was equal in five pairs, lower by one in three, and higher by one in the remaining pair.

Relative to \textsc{Compact-Base}, \textsc{Compact-Objective} had lower median RSS in seven pairs and lower PAR-2 in three; the other pairs favored \textsc{Compact-Base}.

Across all repetitions, runtime ratios favored \textsc{Compact-Objective} only for \textsc{IR} on the two stress sets (Figure~\ref{fig:encoding-effects}(b)). Comparisons within the same execution environment changed some rankings. On the same 47 Stress instances under \textsc{IR-IM-IS}, the median \textsc{Compact-Base}/\textsc{Compact-Objective} runtime ratio was 0.55 using three repetitions and 1.76 using the 47 matching run pairs from repetition~1. Four of the nine comparisons had no run pairs with complete, matching execution settings.

\FloatBarrier
\subsection{Balancing Participant Waiting (RQ2)}
\label{sec:schedule-quality}

We compare refinements of idle-range balance, then examine precedence-stress instances.

\subsubsection{Preserving Idle-Range Balance}

\begin{figure}[!htbp]
\centering
\begingroup%
\expandafter\def\csname QualityOutliersquality-effect0\endcsname{(0,-2) (1,-2) (2,-2) (3,-2) (4,-2) (5,-1) (6,-1) (7,-1) (8,-1) (9,-1) (10,-1)}
\expandafter\def\csname QualityOutliersquality-effect1\endcsname{(0,-698) (1,-534) (2,-498) (3,-493) (4,-327) (5,-317) (6,-312) (7,-301) (8,-299) (9,-298) (10,-217) (11,-208)}
\expandafter\def\csname QualityOutliersquality-effect2\endcsname{(0,-1.58872831311353) (1,-1.31203866721508) (2,-1.28438822547471)}
\expandafter\def\csname QualityOutliersquality-effect3\endcsname{(0,-197)}
\expandafter\def\csname QualityOutliersquality-effect4\endcsname{(0,-663) (1,-494) (2,-461) (3,-440) (4,-315) (5,-305) (6,-288) (7,-287)}
\expandafter\def\csname QualityOutliersquality-effect5\endcsname{(0,-1.0915965900546) (1,-0.880753172831216) (2,0.452853142876054) (3,0.520678236915224) (4,0.65084779614403) (5,0.65084779614403) (6,0.677624645122952) (7,0.713793855509574) (8,0.753400130532854) (9,0.911186914601642) (10,0.911186914601642) (11,0.996302895492097) (12,1.14076527929641) (13,2.56296831681635)}
\expandafter\def\csname QualityOutliersquality-effect6\endcsname{(0,-72) (1,-72) (2,-69) (3,-67) (4,-61) (5,-57) (6,-57) (7,-57) (8,-57) (9,-57) (10,-57) (11,-57) (12,-57) (13,-56) (14,-49) (15,-49) (16,-32) (17,-28) (18,-24) (19,-21)}
\expandafter\def\csname QualityOutliersquality-effect7\endcsname{(0,-55) (1,-54) (2,-53) (3,-53) (4,-51) (5,-51) (6,-46) (7,-44) (8,-35) (9,-30) (10,6) (11,6) (12,7) (13,8) (14,11) (15,11) (16,14) (17,22) (18,27) (19,35) (20,37) (21,37) (22,40) (23,51) (24,53)}
\expandafter\def\csname QualityOutliersquality-effect8\endcsname{(0,0.993130748680711) (1,1.07640147306298) (2,1.29007781576045) (3,1.41407282979979) (4,1.43013918869546) (5,1.44792458050092) (6,1.718796023006) (7,1.78029734880033) (8,1.94723892052965) (9,2.14739565874275) (10,2.64065935135256)}
\pgfplotstableread[col sep=comma,row sep=\\]{%
y,lower_whisker,q1,median,q3,upper_whisker,n,lower,equal,higher\\
8,0,0,0,0,0,84,11,73,0\\
7,-116,-67.5,-35,0,0,84,56,28,0\\
6,-1.06650349329076,-0.479435016093488,-0.10645891645568,0,0,84,45,39,0\\
5,-176,-76.25,-28,0,0,84,57,27,0\\
4,-275,-111.25,-34.5,0,0,84,57,27,0\\
3,-0.55999117277884,-0.27468398148193,0,0.0138879378752694,0.406757355790924,84,30,33,21\\
2,-9,-6.75,0,0,0,84,32,52,0\\
1,0,0,0,2,4,84,10,43,31\\
0,0,0,0,0.371605256006864,0.911186914601642,84,0,43,41\\
}\QualityStats
\def\DrawQualityBox#1#2#3{%
  \pgfplotstablegetelem{#1}{lower_whisker}\of\QualityStats\edef\BoxLo{\pgfplotsretval}
  \pgfplotstablegetelem{#1}{q1}\of\QualityStats\edef\BoxQOne{\pgfplotsretval}
  \pgfplotstablegetelem{#1}{median}\of\QualityStats\edef\BoxMed{\pgfplotsretval}
  \pgfplotstablegetelem{#1}{q3}\of\QualityStats\edef\BoxQThree{\pgfplotsretval}
  \pgfplotstablegetelem{#1}{upper_whisker}\of\QualityStats\edef\BoxHi{\pgfplotsretval}
  \pgfplotstablegetelem{#1}{lower}\of\QualityStats\edef\CountLower{\pgfplotsretval}
  \pgfplotstablegetelem{#1}{equal}\of\QualityStats\edef\CountEqual{\pgfplotsretval}
  \pgfplotstablegetelem{#1}{higher}\of\QualityStats\edef\CountHigher{\pgfplotsretval}
  \edef\BoxData{coordinates {\csname QualityOutliersquality-effect#1\endcsname};}%
  \edef\BoxCommand{\noexpand\addplot[
    boxplot prepared={draw position=#2,lower whisker=\BoxLo,
      lower quartile=\BoxQOne,median=\BoxMed,upper quartile=\BoxQThree,
      upper whisker=\BoxHi},
    boxplot/draw direction=x,boxplot/box extend=.30,boxplot/whisker extend=.22,
    solid,draw=black!75,fill=black!12,line width=.45pt,
    boxplot/every median/.style={draw=black,line width=.9pt},
    mark=o,mark size=.6pt,mark options={draw=black!70,fill=white,line width=.3pt}]
    \BoxData
  }%
  \BoxCommand
  \edef\CountCommand{\noexpand\node[anchor=west,font=\noexpand\sffamily\noexpand\fontsize{8}{9.5}\noexpand\selectfont]
    at (axis cs:#3,#2) {\CountLower/\CountEqual/\CountHigher};}%
  \CountCommand
}%
\begin{tikzpicture}[x=1mm,y=1mm,every node/.style={fig text}]
\path[use as bounding box] (0,0) rectangle (159,113);
\node[fig heading] (panel-a) at (0,110) {(a) Maximum idle ($I_{\max}$)};
\node[fig heading] (panel-b) at (80,110) {(b) Total breaks ($G_\Sigma$)};
\node[fig heading] (panel-c) at (0,55) {(c) Total idle ($I_\Sigma$)};
\node[fig heading] (panel-d) at (80,55) {(d) Idle SD ($\sigma_I$)};
\foreach \xx/\yy in {72/103,152/103,72/48,152/48}{
  \node at (\xx,\yy) {L/E/H};
}

\begin{axis}[figure axis,name=max-axis,at={(18mm,70mm)},anchor=south west,
  width=45mm,height=30mm,scale only axis,
  xmin=-2.25,xmax=.25,ymin=-.6,ymax=.6,
  xtick={-2,-1,0},ytick={0},yticklabels={Lex $-$ IR},
  ytick style={draw=none},xlabel={Difference (slots)},
]
\draw[draw=black!45,dashed,line width=.45pt] (axis cs:0,-.6)--(axis cs:0,.6);
\DrawQualityBox{0}{0}{.42}
\end{axis}

\begin{axis}[figure axis,name=break-axis,at={(98mm,70mm)},anchor=south west,
  width=45mm,height=30mm,scale only axis,
  xmin=-210,xmax=10,ymin=1.3,ymax=3.7,
  xtick={-200,-100,0},ytick={3,2},
  yticklabels={Break $-$ IR,Break $-$ Lex},
  ytick style={draw=none},xlabel={Difference (breaks)},
]
\draw[draw=black!45,dashed,line width=.45pt] (axis cs:0,1.3)--(axis cs:0,3.7);
\DrawQualityBox{3}{3}{25}
\DrawQualityBox{6}{2}{25}
\end{axis}

\begin{axis}[figure axis,name=sum-axis,at={(18mm,15mm)},anchor=south west,
  width=45mm,height=30mm,scale only axis,
  xmin=-750,xmax=100,ymin=1.3,ymax=4.7,
  xtick={-600,-300,0},ytick={4,3,2},
  yticklabels={Lex $-$ IR,Break $-$ IR,Break $-$ Lex},
  ytick style={draw=none},xlabel={Difference (slots)},
]
\draw[draw=black!45,dashed,line width=.45pt] (axis cs:0,1.3)--(axis cs:0,4.7);
\DrawQualityBox{1}{4}{157}
\DrawQualityBox{4}{3}{157}
\DrawQualityBox{7}{2}{157}
\end{axis}

\begin{axis}[figure axis,name=sd-axis,at={(98mm,15mm)},anchor=south west,
  width=45mm,height=30mm,scale only axis,
  xmin=-1.8,xmax=2.9,ymin=1.3,ymax=4.7,
  xtick={-1,0,1,2},ytick={4,3,2},
  yticklabels={Lex $-$ IR,Break $-$ IR,Break $-$ Lex},
  ytick style={draw=none},xlabel={Difference (slots)},
]
\draw[draw=black!45,dashed,line width=.45pt] (axis cs:0,1.3)--(axis cs:0,4.7);
\DrawQualityBox{2}{4}{3.22}
\DrawQualityBox{5}{3}{3.22}
\DrawQualityBox{8}{2}{3.22}
\end{axis}

\coordinate (a-sw) at (max-axis.south west);\coordinate (a-ne) at (max-axis.north east);
\coordinate (b-sw) at (break-axis.south west);\coordinate (b-ne) at (break-axis.north east);
\coordinate (c-sw) at (sum-axis.south west);\coordinate (c-ne) at (sum-axis.north east);
\coordinate (d-sw) at (sd-axis.south west);\coordinate (d-ne) at (sd-axis.north east);
\ReportFigurePanel{paired_schedule_outcomes}{a}{a-sw}{a-ne}{panel-a}
\ReportFigurePanel{paired_schedule_outcomes}{b}{b-sw}{b-ne}{panel-b}
\ReportFigurePanel{paired_schedule_outcomes}{c}{c-sw}{c-ne}{panel-c}
\ReportFigurePanel{paired_schedule_outcomes}{d}{d-sw}{d-ne}{panel-d}
\end{tikzpicture}%
\endgroup%
\caption{Paired differences between selected schedules on \MainQualityN{} instances solved to optimality under all three objectives. Lex denotes \textsc{IR-IM-IS}; Break denotes \textsc{BG-d$\infty$}. Panels show maximum idle, total breaks, total idle, and idle SD, each on its own scale. Differences are left minus right; dashed lines mark zero. Boxes show medians and IQR; whiskers reach the most extreme values within 1.5 IQR of the box. Circles mark outliers; L/E/H counts lower/equal/higher values under the left objective.}
\label{fig:schedule-quality-outcomes}
\Description{Four panels separate the waiting measures. Maximum idle compares Lex minus IR; total breaks compares Break minus IR and Break minus Lex; total idle and participant-idle standard deviation each show all three comparisons. Every row uses 84 pairs of selected schedules, one pair per instance, and reports lower, equal, and higher counts. All axes are linear and have their own labeled units.}
\end{figure}
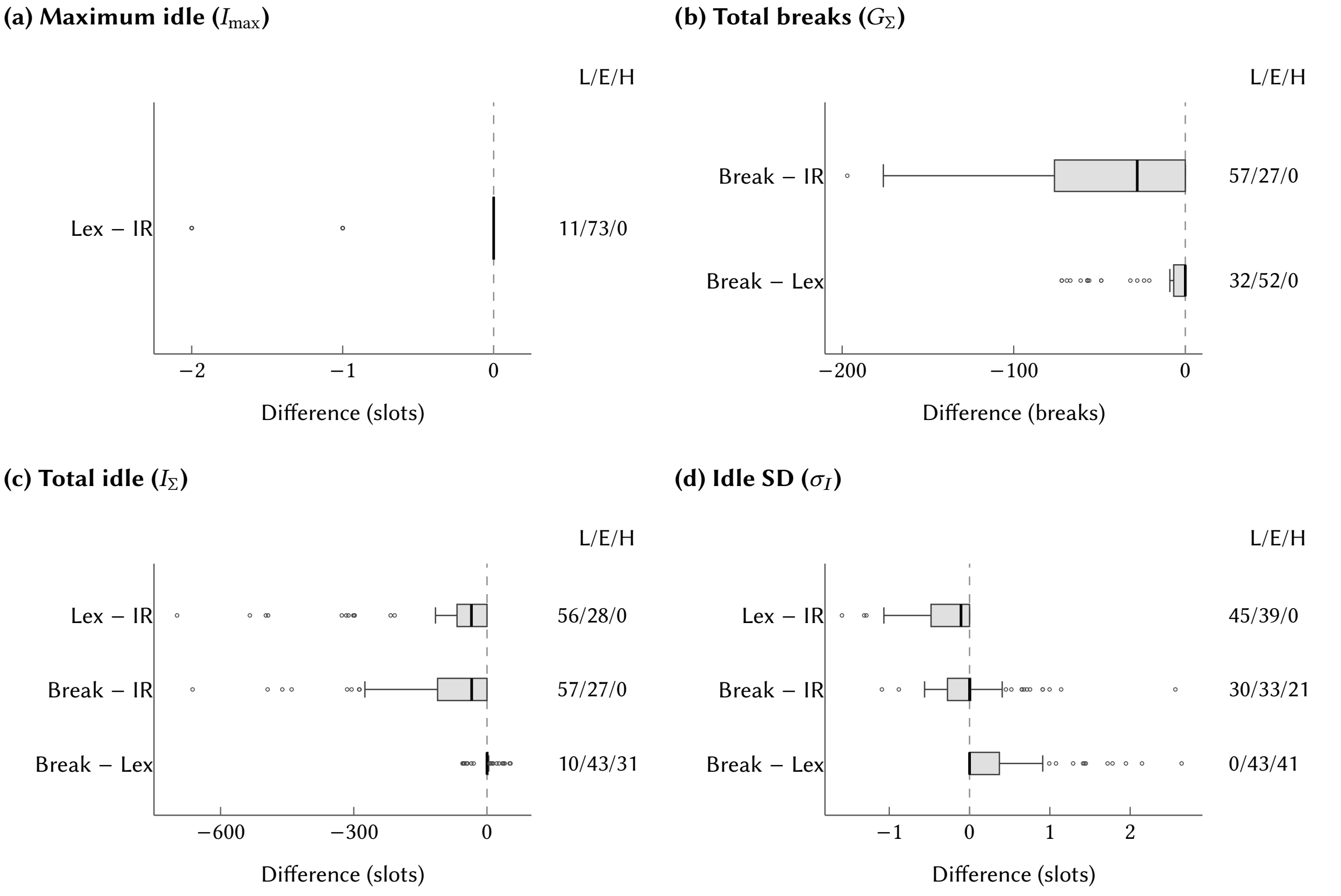

On the \MainQualityN{} All-126 instances solved to optimality under \textsc{BG-d$\infty$}, \textsc{IR}, and \textsc{IR-IM-IS}, the selected \textsc{IR-IM-IS} schedules retained the optimal range of \textsc{IR}. Maximum idle decreased on \MainQualityMaximumLower{} instances and total idle on \MainQualityTotalLower{}; neither increased (Figure~\ref{fig:schedule-quality-outcomes}). Table~\ref{tab:main-quality-medians} summarizes quality medians.

\FloatBarrier

\begin{table}[htbp]
\centering
\caption{Median quality of selected schedules on \MainQualityN{} instances solved to optimality under all three objectives. Bold numerical values mark the lowest displayed median in each column, including rounding ties.}
\label{tab:main-quality-medians}
\begin{tabular*}{0.86\textwidth}{@{\extracolsep{\fill}}lrrrrr@{}}
\toprule
Objective & $G_\Sigma$ & $\Delta_I$ & $I_{\max}$ & $I_\Sigma$ & $\sigma_I$ \\
\midrule
\MainQualityMedianRows
\bottomrule
\end{tabular*}
\Description{Schedule-quality measures for three objectives on instances solved to optimality under all three: median break count, idle range as the primary balance measure, maximum idle, total idle, and population standard deviation of participant idle totals.}
\end{table}

\FloatBarrier

The medians in Table~\ref{tab:main-quality-medians} summarize each objective separately; Figure~\ref{fig:schedule-quality-outcomes} shows differences between objectives on the same instance. Relative to \textsc{IR}, \textsc{IR-IM-IS} also had lower $\sigma_I$ on 45 instances and equal $\sigma_I$ on 39.

Compared with \textsc{IR-IM-IS}, \textsc{BG-d$\infty$} returned fewer breaks on 32 instances. Its total idle was lower on 10, equal on 43, and higher on 31; its $\sigma_I$ was higher on 41 and equal on 43.

\FloatBarrier

To examine the maximum-idle priority separately, we compared \textsc{IR-IS} and \textsc{IR-IM-IS} on the \PriorityQualityN{} instances from this All-126 set that were also solved to optimality under \textsc{IR-IS}. The selected \textsc{IR-IS} and \textsc{IR-IM-IS} schedules had equal idle range, maximum idle, and total idle on every instance in this common set. Thus, the maximum-idle tier provided no additional improvement in these three measures on this subset.

\FloatBarrier
\subsubsection{Total versus Squared Idle after Balancing}
\label{sec:squared-idle-results}

We compared \textsc{IR-IM-IS} and \textsc{IR-IM-ISQ} to examine the last criterion after minimizing idle range and maximum idle. Each objective reached optimality on 84 of the 126 instances, with 83 instances in common. Each solved one instance on which the other timed out.

Table~\ref{tab:final-tier-quality} shows paired differences, defined as \textsc{IR-IM-ISQ} minus \textsc{IR-IM-IS}. Lower, Equal, and Higher count negative, zero, and positive differences; Min and Max give the smallest and largest differences.

\begin{table}[!htbp]
\centering
\caption{Paired differences between selected \textsc{IR-IM-ISQ} and \textsc{IR-IM-IS} schedules on 83 instances solved to optimality under both objectives.}
\label{tab:final-tier-quality}
\begin{tabular*}{0.86\textwidth}{@{\extracolsep{\fill}}lrrrrr@{}}
\toprule
Measure & Lower & Equal & Higher & Min & Max \\
\midrule
$I_{\mathrm{SQ}}$ (slots$^2$) & 15 & 68 & 0 & $-143$ & 0 \\
$\sigma_I$ (slots) & 15 & 68 & 0 & $-0.520$ & 0 \\
$I_\Sigma$ (slots) & 0 & 69 & 14 & 0 & $+12$ \\
\bottomrule
\end{tabular*}
\Description{Differences are IR-IM-ISQ minus IR-IM-IS on 83 instances. Squared idle and participant-idle SD decrease on 15 instances and are equal on 68; total idle is higher on 14 and equal on 69. The last two columns give minimum and maximum differences in the stated units.}
\end{table}

On all 83 instances, the selected schedules had equal idle range and maximum idle. \textsc{IR-IM-ISQ} had lower squared idle and SD on 15 instances and equal values on 68. Total idle increased on 14 instances, all among those 15, and was unchanged on the remaining 69. Thus, squared idle and SD can decrease at the expense of higher total waiting while preserving the range and maximum-idle priorities.

Appendix~Table~\ref{tab:waiting-objective-comparison} compares additional waiting objectives, including standalone \textsc{ISQ}.

\FloatBarrier
\Needspace{7\baselineskip}
\subsubsection{Precedence-Stress Instances}

On Stress, \StressQualityN{} instances were solved to optimality under both \textsc{IR} and \textsc{IR-IM-IS}. The selected \textsc{IR-IM-IS} schedules preserved idle range on all \StressQualityN{}; maximum idle decreased on 4 and total idle on \StressQualityTotalLower{}. On Stress-high, \StressHighQualityN{} instances were solved to optimality under both objectives. Idle range and maximum idle were unchanged in their selected schedules, while total idle decreased on \StressHighQualityTotalLower{}. All remaining paired values for these measures were equal (Appendix~Table~\ref{tab:app-stress-quality}).

\FloatBarrier

\Needspace{9\baselineskip}
\subsection{Ablation Study (RQ3)}

Each of the 126 benchmark instances contains at least one participant requesting at most one meeting, allowing individual break bounds. For \textsc{BG-d2}, this zero-break simplification (ZB) gave a median per-instance clause reduction of 13.33\% relative to \textsc{Compact-Base}. Omitting unused objective states and reusing equivalent states (DS), combined with ZB, gave a 16.77\% reduction against the same baseline. Adding adaptive participant at-most-one encoding (AMO) reduced clauses by a further 3.93\% relative to that combined configuration (Table~\ref{tab:main-ablation-primary}).

\begin{table}[!htbp]
\centering
\caption{Component ablation on 126 instances. Counts and paired clause reductions are medians; bold numerical values mark the smallest displayed count within each objective, including rounding ties.}
\label{tab:main-ablation-primary}
\begingroup
\setlength{\tabcolsep}{4pt}
\begin{tabular*}{\textwidth}{@{\extracolsep{\fill}}l*{6}{c}rrr@{\hspace{1em}}r@{}}
\toprule
 & \multicolumn{6}{c}{Enabled components} & & & & \\
\cmidrule(lr){2-7}
Objective & B & ZB & DS & RC & AMO & AV & Variables & Clauses & Literals & \shortstack{Clause reduction\\(\%)} \\
\midrule
\textsc{BG-d2} & \tablecheck &  &  &  &  &  & 31{,}250.5 & 81{,}521.5 & 197{,}520.5 & -- \\
 & \tablecheck & \tablecheck &  &  &  &  & 27{,}176.5 & 72{,}991.5 & 167{,}541.5 & 13.33 \\
 & \tablecheck &  & \tablecheck &  &  &  & 30{,}589 & 79{,}668.5 & 193{,}283.5 & 2.20 \\
 & \tablecheck & \tablecheck & \tablecheck &  &  &  & \textbf{26{,}515} & 71{,}138.5 & 163{,}304.5 & 16.77 \\
 & \tablecheck & \tablecheck & \tablecheck &  & \tablecheck &  & 27{,}736 & \textbf{66{,}318} & \textbf{153{,}675} & 3.93$^\dagger$ \\
\addlinespace
\textsc{IR} & \tablecheck &  &  &  &  &  & 26{,}727.5 & 72{,}487 & 167{,}045.5 & -- \\
 & \tablecheck &  & \tablecheck &  &  &  & 26{,}535.5 & 72{,}109 & 166{,}289.5 & 0.75 \\
 & \tablecheck &  &  & \tablecheck &  &  & 26{,}720 & 72{,}464.5 & 167{,}000.5 & 0.04 \\
 & \tablecheck &  & \tablecheck & \tablecheck &  &  & \textbf{26{,}528} & 72{,}086.5 & 166{,}244.5 & 0.80 \\
 & \tablecheck &  & \tablecheck & \tablecheck & \tablecheck &  & 27{,}750 & 67{,}255.5 & 156{,}592 & 3.69$^\dagger$ \\
 & \tablecheck &  & \tablecheck & \tablecheck & \tablecheck & \tablecheck & 27{,}739 & \textbf{67{,}201.5} & \textbf{156{,}464.5} & 0.19$^\dagger$ \\
\addlinespace
\textsc{IR-IM-IS} & \tablecheck &  &  &  &  &  & 26{,}727.5 & 72{,}612.5 & 167{,}171 & -- \\
 & \tablecheck &  & \tablecheck &  &  &  & 26{,}535.5 & 72{,}234.5 & 166{,}415 & 0.75 \\
 & \tablecheck &  &  & \tablecheck &  &  & 26{,}720 & 72{,}590 & 167{,}126 & 0.04 \\
 & \tablecheck &  & \tablecheck & \tablecheck &  &  & \textbf{26{,}528} & 72{,}212 & 166{,}370 & 0.79 \\
 & \tablecheck &  & \tablecheck & \tablecheck & \tablecheck &  & 27{,}750 & 67{,}381 & 156{,}717.5 & 3.68$^\dagger$ \\
 & \tablecheck &  & \tablecheck & \tablecheck & \tablecheck & \tablecheck & 27{,}739 & \textbf{67{,}326} & \textbf{156{,}589} & 0.20$^\dagger$ \\
\bottomrule
\end{tabular*}
\endgroup
\par\noindent\parbox{\textwidth}{\textit{Note:} B is \textsc{Compact-Base} \textbf{(ours)}; ZB, zero-break simplification; DS, demand-driven states; RC, direct range costs; AMO, adaptive at-most-one encoding; and AV, attainable-value compression. Percentages are relative to B unless marked $\dagger$, in which case they are relative to the preceding row. Rows with AMO but without AV are \textsc{Compact-Objective} \textbf{(ours)}. AV rows are optional extensions.}
\Description{A component matrix with median variables, clauses, literal occurrences, and paired clause reductions under BG-d2, IR, and IR-IM-IS.}
\end{table}

\FloatBarrier

For both idle objectives, DS and direct soft-clause penalties for range (RC) removed less than 1\% of clauses relative to \textsc{Compact-Base}. Adding AMO to the combined DS and RC configuration reduced clauses by a further 3.69\% for \textsc{IR} and 3.68\% for \textsc{IR-IM-IS}. The AMO encoding increased the median variable count while reducing clauses and literals. Optional attainable-value compression gave a further reduction of 0.20\% or less.

In the objective-encoding ablation, direct idle-slot clauses for \textsc{IS} used a median of 5.92\% fewer clauses per instance than dense thresholds. Relative to dense thresholds, attainable-value compression reduced clauses by at most 0.20\% for the other objectives (Appendix~Table~\ref{tab:app-objective-layer-ablation}).

\Needspace{8\baselineskip}
\section{Discussion}
\label{sec:discussion}

Different schedules can have the same minimum idle range but different total waiting times. The lexicographic refinement selects among them by minimizing maximum idle, then total idle among schedules tied on both preceding criteria. This keeps balance as the primary preference while using waiting duration to distinguish equally balanced schedules. The priority order makes explicit the trade-off between balance and total waiting discussed in equity optimization \parencite{karsu2015,matl2018}.

On the additional common set, \textsc{IR-IS} attained the same range, maximum idle, and total idle as \textsc{IR-IM-IS}. The maximum-idle tier expresses a preference for limiting the largest individual idle total before aggregate waiting. Its benefit on instances where maximum and total idle compete remains an open empirical question.

Break counts, squared idle, and idle range express different scheduling preferences. Break counts measure interruptions; squared idle penalizes large individual idle totals, while range measures the difference between the largest and smallest totals. \textsc{IR-IM-ISQ} retains the range and maximum-idle priorities of \textsc{IR-IM-IS} and changes only the last criterion. Their comparison shows that reducing squared idle and participant-idle SD can increase total waiting. Standalone \textsc{ISQ} gives no prior preference to range or maximum idle (Appendix~Table~\ref{tab:waiting-objective-comparison}). SD describes variation across participants within a selected schedule. The balance comparison uses absolute idle totals among participants with at least two meetings, without adjustment for meeting load.

Compact encoding addresses the computational cost of representing these preferences. Relative to the earlier \textsc{Compact-Base} encoding \parencite{nguyen2026compact}, \textsc{Compact-Objective} reduced clauses and literal occurrences, sometimes with more auxiliary variables. It had lower median memory use than \textsc{Direct}, while runtime comparisons with \textsc{Compact-Base} varied by objective and data set. Thus, additional clause savings did not consistently translate into shorter runtimes; formula size, memory, and runtime need separate assessment.

Comparisons with a fixed solver build and computing environment would help isolate the runtime effect of the encodings. The evaluation covers variants of 20 base instances. Events with more varied meeting loads would test how prioritizing idle range affects participants with different numbers of meetings.

\section{Conclusion}
\label{sec:conclusions}

We study balance-preserving refinements in B2B meeting scheduling by distinguishing schedules with the same minimum idle range. The three-tier objective prioritizes maximum idle, then total idle among schedules tied on both earlier criteria. Exact MaxSAT encodings preserve feasible schedules, objective costs, and the priority order. On \MainQualityN{} benchmark instances solved to optimality under the break, range, and lexicographic objectives, the selected lexicographic schedules retained the optimal range. Relative to the selected range-only schedules, maximum idle decreased on \MainQualityMaximumLower{} instances and total idle on \MainQualityTotalLower{}. On the \PriorityQualityN{}-instance subset, range followed directly by total idle matched the three-tier criterion values.

\textsc{Compact-Base} used the feasibility encoding from earlier work and required fewer clauses than \textsc{Direct}, our independent implementation based on the 2022 MaxSAT model. \textsc{Compact-Objective} extended this encoding and reduced clauses and literal occurrences further, sometimes with more auxiliary variables. It had lower median peak memory usage than \textsc{Direct} in all nine objective--data-set comparisons. Runtime rankings changed in comparisons within the same execution environment. The empirical findings concern variants of 20 base instances and the tested software environments.

\Needspace{10\baselineskip}
\section{Code and Data Availability}
\label{sec:code-data-availability}
Source code, benchmark instances, individual run results, and reproduction instructions are available at \url{https://github.com/Longnguyen1555/Business-Meeting-Scheduling-Problem} (\href{https://github.com/Longnguyen1555/Business-Meeting-Scheduling-Problem/tree/7dc6e593bd85c11dbbfd887a4f550ea937b6710d}{submission version}). The README describes how to reproduce the reported analyses. Source code is licensed under the MIT License. Author-generated experimental data, analysis outputs, and records of instance generation are licensed under CC BY 4.0. Benchmark contents, including those retained in the derived Stress and Stress-high instances, remain subject to their original terms \parencite{bofill2022}.

\printbibliography

\appendix

\begin{samepage}
\section{Correctness Proofs}
\label{app:encoding-details}

This appendix proves the encoding results stated in Section~\ref{sec:compact-framework}.

\subsection{Feasibility Preservation}
\label{app:filtering-preservation}

Algorithm~\ref{alg:domain-filtering} repeats precedence, participant non-overlap,
and capacity filtering until no domain changes. For participant $p$, a bipartite
graph connects meetings to allowed slots: it has an edge $(m,t)$ for each
$m\in M_p$ and $t\in D_m$. A matching covering $M_p$ assigns all of that
participant's meetings to distinct allowed slots. Capacity filtering uses
singleton current domains, including those obtained by earlier filtering steps.
\end{samepage}

\begin{algorithm}[htbp]
\caption{Reduce meeting domains before CNF generation}
\label{alg:domain-filtering}
\begin{algorithmic}[1]
\Require initial domains $(D_m^0)_{m\in M}$, precedence arcs $E$, participant sets $(M_p)_{p\in P}$, and capacity $K$
\Ensure reduced domains $(D_m)_{m\in M}$ or a report that the instance is infeasible
\State $D_m\gets D_m^0$ for every meeting $m$
\Repeat
  \State $D^{\mathrm{old}}\gets D$
  \ForAll{$(i,j)\in E$}
    \State remove $t\in D_i$ if no $u\in D_j$ satisfies $t<u$
    \State remove $u\in D_j$ if no $t\in D_i$ satisfies $t<u$
  \EndFor
  \ForAll{$p\in P$}
    \State remove $t$ from $D_m$ if edge $(m,t)$ belongs to no matching covering $M_p$
  \EndFor
  \ForAll{$t\in T$}
    \State $F_t\gets\{m\in M:D_m=\{t\}\}$
    \If{$|F_t|>K$}
      \State \Return infeasible
    \ElsIf{$|F_t|=K$}
      \State remove $t$ from $D_m$ for every $m\notin F_t$
    \EndIf
  \EndFor
  \If{some domain is empty or a participant has no covering matching}
    \State \Return infeasible
  \EndIf
\Until{$D=D^{\mathrm{old}}$}
\State \Return $D$
\end{algorithmic}
\end{algorithm}

\begin{proof}[Proof of Theorem~\ref{thm:domain-preservation}]
Every iteration that changes a domain removes at least one value from the
finite initial domains; an iteration with no change ends the algorithm.
Fix a feasible schedule $S$. Initially, $S(m)\in D_m^0$ for every meeting.
Suppose all selected values remain in the current domains. The values $S(i)$
and $S(j)$ support one another on every precedence arc. For each participant
$p$, the edges $\{(m,S(m)):m\in M_p\}$ form a matching covering $M_p$.
If $K$ meetings have domain $\{t\}$, they occupy $t$ in $S$, so capacity
excludes $t$ for every other meeting. Each rule therefore preserves the
invariant $S(m)\in D_m$ through termination. An empty domain, absence of a
covering matching, or more than $K$ meetings fixed at one slot would contradict
this invariant and feasibility of $S$. Each infeasibility report is therefore
sound.
\end{proof}

\FloatBarrier
\subsection{Shared Suffix Encoding}
\label{app:precedence-preservation}

For a precedence relation $(i,j,\ell)$, let $\ell$ denote its minimum slot
separation; the strict precedence constraint (F4) has $\ell=1$. Sort the slots
with assignment variables as $A_i=(a_{i,0},\ldots,a_{i,n_i-1})$, using zero-based positions.
For a slot $u\in A_j$ of successor meeting $j$, let $c(i,j,u)$ be the first position whose
value exceeds $u-\ell$, or $n_i$ if no such position exists. The assignments
from that position onward are exactly the predecessor values incompatible
with $u$. Define
\begin{equation}
 q_{i,c}\leftrightarrow
 \bigvee_{r=c}^{n_i-1}x_{i,a_{i,r}}.
\label{eq:shared-suffix-definition}
\end{equation}
The encoding adds
\begin{equation}
 x_{j,u}\rightarrow\neg q_{i,c(i,j,u)}.
\label{eq:shared-suffix-link}
\end{equation}
The suffix literal represents one forbidden suffix and is reused by all
successor slot choices with the same cut position. At $c=0$, exactly-one assignment makes
the suffix true, so the link becomes $\neg x_{j,u}$. At $c=n_i$, the suffix is
false and no link is needed.

Only cut positions used by precedence constraints and lying strictly between
zero and $n_i$ require suffix states. If these positions are
$c_1<\cdots<c_s$, define the last state by
Equation~\eqref{eq:shared-suffix-definition}, then work backward using
\begin{equation}
 q_{i,c_r}\leftrightarrow
 \left(\bigvee_{v=c_r}^{c_{r+1}-1}x_{i,a_{i,v}}\right)
 \lor q_{i,c_{r+1}}
 \qquad(r<s).
\label{eq:shared-suffix-recurrence}
\end{equation}
A final suffix containing one assignment literal reuses that literal. All other
equivalences are encoded in both directions as hard clauses.

\begin{proof}[Proof of Proposition~\ref{prop:precedence-equivalence}]
Sort $A_i=(a_{i,0},\ldots,a_{i,n_i-1})$. By the definition of the cut,
$a_{i,r}+\ell\le u$ for $r<c(i,j,u)$, while
$a_{i,r}+\ell>u$ for $r\ge c(i,j,u)$. Exactness of the suffix state makes
Equation~\eqref{eq:shared-suffix-link} equivalent, after auxiliary variables are
ignored, to the conjunction of the pairwise conflict clauses
$\neg x_{i,a_{i,r}}\lor\neg x_{j,u}$ with $r\ge c(i,j,u)$.
The boundary cuts exclude either all or none of the predecessor values and give
the same projected restriction. Repeating the argument for every relation and
successor value proves equality of the two projections.

Every direct precedence arc appears in its longest-path closure. Conversely,
summing the unit strict-precedence inequalities along a path of length $\ell$
gives $S(i)+\ell\le S(j)$. The direct arcs and their closure therefore admit
the same schedules. The closure is computed on an acyclic precedence graph;
a directed cycle of strict precedence already makes the instance infeasible.
\end{proof}

\subsection{Objective Costs and Weights}
\label{app:objective-identities}

The following definitions specify the auxiliary variables and soft costs used
by Algorithm~\ref{alg:compile-wcnf}. Boolean equivalences are hard constraints.
For example, $z\leftrightarrow\bigvee_i a_i$ gives
$\neg a_i\lor z$ for each $i$ and $\neg z\lor\bigvee_i a_i$.
The corresponding conjunction definition gives $\neg z\lor a_i$ for each $i$
and $z\lor\bigvee_i\neg a_i$. Empty disjunctions are false.

\paragraph{Occupancy, span and thresholds.}
Occupancy is fixed by Equation~\eqref{eq:occupancy-channeling}. Define the
inclusive prefix and suffix by
\begin{align}
 \mathit{pre}_{p,t}&\leftrightarrow
   \mathit{pre}_{p,t-1}\lor y_{p,t},&
 \mathit{pre}_{p,0}&=0, \label{eq:prefix-definition}\\
 \mathit{suf}_{p,t}&\leftrightarrow
   y_{p,t}\lor\mathit{suf}_{p,t+1},&
 \mathit{suf}_{p,h+1}&=0. \label{eq:suffix-definition}
\end{align}
Suffixes beyond $h$ are false. At slots where no meeting is possible,
neighboring prefix/suffix states share a literal. Idle counting still includes
all event slots between the first and last meetings.

For $p\in P^\star$, the first-meeting indicator and idle thresholds satisfy
\begin{align}
 f_{p,t}&\leftrightarrow
   y_{p,t}\land\neg\mathit{pre}_{p,t-1},
   \label{eq:first-meeting-definition}\\
 f_{p,t}&\rightarrow
   \left(\theta_{p,k}\leftrightarrow
   \mathit{suf}_{p,t+r_p+k-1}\right).
   \label{eq:span-threshold-definition}
\end{align}
The second relation is imposed for every possible first slot $t$ and every
retained threshold level $k$. Its two CNF clauses are
$\neg f_{p,t}\lor\neg\theta_{p,k}\lor\mathit{suf}_{p,t+r_p+k-1}$
and $\neg f_{p,t}\lor\theta_{p,k}\lor\neg\mathit{suf}_{p,t+r_p+k-1}$,
with constants simplified.

Meeting assignment and participant non-overlap give each $p\in P^\star$
exactly $r_p$ occupied slots and a unique first meeting. If it occurs at $t$
and the last meeting occurs at $b$, then
$I_p=b-t+1-r_p$. Thus $I_p\ge k$ holds exactly when
$b\ge t+r_p+k-1$, which is the suffix condition in
Equation~\eqref{eq:span-threshold-definition}. The hard clauses therefore fix
each retained threshold to $[I_p\ge k]$. For break objectives, exact
cardinality thresholds instead define
$\beta_{p,k}\leftrightarrow[\sum_t e_{p,t}\ge k]$ for
$1\le k\le\max\{0,r_p-1\}$.

\paragraph{Maximum and range.}
Let $Q_p$ be a nonnegative integer count over a nonempty participant set $R$,
and let $\eta_{p,k}\leftrightarrow[Q_p\ge k]$ be its exact thresholds.
Thresholds above a participant's upper bound are false. Up to a common upper
bound $U$, define
\begin{equation}
 m_k\leftrightarrow\bigvee_{p\in R}\eta_{p,k},\qquad
 n_k\leftrightarrow\bigwedge_{p\in R}\eta_{p,k},\qquad
 \rho_k\leftrightarrow m_k\land\neg n_k.
\label{eq:count-extremes-definition}
\end{equation}
The idle and break counts use separate copies of these indicators: idle uses
$R=P^\star$ and $\eta=\theta$; breaks use $R=P$ and $\eta=\beta$. Counting levels gives
\begin{equation}
 \max_{p\in R}Q_p=\sum_{k=1}^{U}m_k,\quad
 \min_{p\in R}Q_p=\sum_{k=1}^{U}n_k,\quad
 \max_{p\in R}Q_p-\min_{p\in R}Q_p=\sum_{k=1}^{U}\rho_k.
\label{eq:count-extremes-cost}
\end{equation}
Consequently, single-literal soft clauses $\neg m_k$ charge the maximum, and
$\neg m_k\lor n_k$ charge the range, each with weight one.
The range clauses are violated under exactly the same assignments as $\neg\rho_k$ and avoid
creating $\rho_k$ when it has no other use. General \textsc{BG-d2} retains
the hard range indicators and enforces $\sum_k\rho_k\le2$.
If $P^\star$ is empty, all idle costs are zero and these idle states are
omitted. For a single participant, the range is zero.

\begin{proof}[Proof of Lemma~\ref{lem:break-end-counting}]
Take a maximal internal empty block $[a,b]$. It has an occupied slot before
$a$, an empty slot $b$, and an occupied slot $b+1$. Equation~\eqref{eq:break-end}
therefore makes $e_{p,b}$ true. No other slot in the block is followed by
occupancy. Conversely, a true $e_{p,t}$ identifies a block of one or more empty
slots ending at $t$, with occupancy on both sides. Extending the block maximally to the left
recovers one internal block. The two mappings are inverse, and summing the
indicators gives Equation~\eqref{eq:break-end-counting}.
\end{proof}

\begin{proof}[Proof of Proposition~\ref{prop:zero-break-simplification}]
Nonnegative break counts and $G_{p_0}=0$ imply $\min_pG_p=0$. Hence
$\Delta_G=\max_pG_p$, so $\Delta_G\le2$ holds exactly when every $G_p\le2$.
Lemma~\ref{lem:break-end-counting} replaces each $G_p$ by its break-end sum.
\end{proof}

\begin{proof}[Proof of Lemma~\ref{lem:direct-idle-cost}]
The clause in Equation~\eqref{eq:direct-idle-clause} is false exactly when an
earlier meeting exists, the current slot is empty, and a later meeting exists.
These conditions define an internal idle slot. Summing one violation over all
such participant--slot pairs gives $I_\Sigma$.
\end{proof}

\paragraph{Attainable-value costs.}
For the threshold objectives, use $k$ for a dense integer level and
$v_{p,i}$ for a retained attainable level. Let
$V_p=(v_{p,0},\ldots,v_{p,a_p})$ be the locally attainable idle values of
participant $p$. Test each candidate first/last slot pair $(a,b)$ for a
matching covering $M_p$ within $[a,b]$ and using both endpoints.
Every successful pair contributes $b-a+1-r_p$ to $V_p$.
Every globally feasible agenda passes this local test, so $I_p\in V_p$.

The minimum $v_{p,0}$ may be positive. At a dense level $k$, the condition
$[I_p\ge k]$ is constant true if $k\le v_{p,0}$, constant false if
$k>v_{p,a_p}$, and otherwise equals the retained threshold at the smallest
attainable value at least $k$. Consecutive dense levels can share a range or
maximum clause when, for each participant, the threshold condition uses the
same retained literal or Boolean constant throughout the group. Their
$m_k,n_k,\rho_k$ values are then identical, and the clause weight is the number
of represented levels.

Dense total and squared-idle costs use single-literal soft clauses $\neg\theta_{p,k}$ with
weights $1$ and $2k-1$, respectively. Compressed costs use single-literal soft clauses
$\neg\theta_{p,v_{p,i}}$ with the weights in
Equations~\eqref{eq:compressed-idle-total} and~\eqref{eq:compressed-idle-square},
and include their base constants. For a positive
constant cost $c$, introduce a new variable $g$. Add $g$ as a hard clause and
$\neg g$ as a soft clause of weight $c$. The hard clause forces the soft clause
to be violated, contributing the constant cost. Constant contributions from
compressed range or maximum levels use the same construction.

\begin{proof}[Proof of Lemma~\ref{lem:threshold-counting}]
In the dense representation, exactly the variables
$\theta_{p,1},\ldots,\theta_{p,I_p}$ are true. Their sum is therefore $I_p$.
The identity $v^2=\sum_{k=1}^{v}(2k-1)$ similarly gives $I_p^2$ and hence
Equation~\eqref{eq:squared-idle-unary}.
If $I_p=v_{p,j}$, exactly the retained thresholds with $1\le i\le j$ are
true, so the weighted sums telescope:
\begin{align}
 v_{p,0}+
 \sum_{i=1}^{a_p}(v_{p,i}-v_{p,i-1})
 \theta_{p,v_{p,i}} &= I_p,\label{eq:compressed-idle-total}\\
 v_{p,0}^2+
 \sum_{i=1}^{a_p}(v_{p,i}^2-v_{p,i-1}^2)
 \theta_{p,v_{p,i}} &= I_p^2.\label{eq:compressed-idle-square}
\end{align}
Since the grouped range and maximum levels have identical threshold conditions,
weighting each clause by the number of dense levels in its group preserves
their costs, including constant contributions. For example,
$V_p=(2,4)$ gives $I_p=2+2\theta_{p,4}$ and
$I_p^2=4+12\theta_{p,4}$.
\end{proof}

\begin{proof}[Proof of Proposition~\ref{prop:lexicographic-weights}]
By construction, $I_{\max}\le U_{\max}$, $I_\Sigma\le U_\Sigma$, and
$I_{\mathrm{SQ}}\le U_{\mathrm{SQ}}$ for every feasible schedule. Let $U_L$
bound the last tier. In a two-tier objective, the lower-tier
difference between two schedules lies in $[-U_L,U_L]$. A one-unit improvement
in the first tier reduces that tier's contribution by $U_L+1$, which exceeds
any opposing change in the lower tier.

For a three-tier objective, the maximum contribution of the last two tiers is
$U_{\max}(U_L+1)+U_L=(U_{\max}+1)(U_L+1)-1$. The range weight in
Equation~\eqref{eq:three-tier-weighted-cost} exceeds this value. The weight assigned
to maximum idle, $U_L+1$, similarly dominates the last tier. The scalar and lexicographic
orders are therefore identical.
\end{proof}

\subsection{Objective Preservation}
\label{app:framework-correctness}

\begin{proof}[Proof of Theorem~\ref{prop:objective-preservation}]
In the compact construction, each assignment satisfying the hard clauses selects
one slot from $D_m\subseteq D_m^0$ for every meeting, so (F1) holds. The non-overlap,
capacity and precedence clauses impose (F2)--(F4); pairwise and shared-suffix
precedence clauses exclude the same incompatible assignments. For
\textsc{BG-d2}, exact range indicators or the equivalent individual bounds
from Proposition~\ref{prop:zero-break-simplification} also enforce
$\Delta_G\le2$. Every decoded schedule therefore belongs to $\mathcal F_q$.

Conversely, domain preservation retains the selected slots of every schedule
in $\mathcal F_q$. Its assignment literals satisfy the assignment, capacity
and precedence constraints. In a commander at-most-one group, pairwise clauses
within each subgroup and recursive constraints on commanders exclude two true
inputs. Every valid input assignment extends by setting each commander to
the disjunction of its subgroup's inputs, recursively. The occupancy, prefix,
suffix and threshold variables take their defined values, and any required
\textsc{BG-d2} bounds hold because $\Delta_G\le2$.

Break-end counting, direct idle clauses, and dense or compressed threshold
identities give the exact component costs, including constant terms. The soft
cost equals the stated single-criterion cost or weighted lexicographic cost.
The dominating weights preserve the lexicographic order, so projecting optimal
weighted-CNF assignments gives exactly the optimal schedule set, with the same
objective vectors.
\end{proof}

\Needspace{6\baselineskip}
\section{Supporting Results}
\label{app:supporting-results}

The following tables compare additional waiting objectives and schedule quality on the precedence-stress sets, followed by additional encoding-size comparisons.

\subsection{Additional Schedule-Quality Results}
\label{app:extended-quality}

To examine penalties on long waits, we compared \textsc{BG-d$\infty$}, \textsc{IS}, \textsc{ISQ}, \textsc{IM-IS}, and \textsc{IR-IS}. Results for the four idle objectives cover 95 of the 126 benchmark instances; the other 31 have no run records in this experiment. Of the 95 instances, 23 were infeasible and \textsc{ISQ} timed out on 10. Table~\ref{tab:waiting-objective-comparison} therefore compares the 62 instances solved to optimality under all five objectives.

\begin{table}[!htbp]
\centering
\caption{Median quality of selected schedules on 62 instances solved to optimality under all five objectives. Bold values mark the lowest displayed median in each column, including rounding ties.}
\label{tab:waiting-objective-comparison}
\begin{tabular*}{0.86\textwidth}{@{\extracolsep{\fill}}lrrrrr@{}}
\toprule
Objective & $G_\Sigma$ & $\Delta_I$ & $I_{\max}$ & $I_\Sigma$ & $I_{\mathrm{SQ}}$ \\
\midrule
\textsc{BG-d$\infty$} & \textbf{1.0} & 2.0 & 2.0 & 5.0 & 16.0 \\
\textsc{IS} & 3.0 & 2.0 & 2.0 & \textbf{3.5} & 5.0 \\
\textsc{ISQ} & 3.0 & \textbf{1.5} & \textbf{1.5} & \textbf{3.5} & \textbf{4.5} \\
\textsc{IM-IS} & 3.0 & \textbf{1.5} & \textbf{1.5} & \textbf{3.5} & \textbf{4.5} \\
\textsc{IR-IS} & 4.0 & \textbf{1.5} & \textbf{1.5} & 8.0 & 16.0 \\
\bottomrule
\end{tabular*}
\Description{Five waiting objectives are compared on the same 62 instances: break minimization, total idle, squared idle, maximum then total idle, and range then total idle. Columns show medians of total breaks, idle range, maximum idle, total idle, and squared idle. Every tied column minimum is highlighted.}
\end{table}

\textsc{ISQ} and \textsc{IS} both had median total idle 3.5, with median squared idle 4.5 and 5.0, respectively. In paired schedules, \textsc{ISQ} reduced squared idle on 14 instances and increased total idle on 10. Compared with \textsc{IR-IS}, its idle range was higher on 12 instances and equal on the other 50. These comparisons distinguish penalizing large idle totals from giving idle range first priority.

Table~\ref{tab:app-stress-quality} compares selected \textsc{IR} and \textsc{IR-IM-IS}
schedules on Stress and Stress-high, restricting each set to instances solved
to optimality under both objectives.

\begin{table}[!htbp]
\centering
\caption{Changes from selected \textsc{IR} to \textsc{IR-IM-IS} schedules on instances solved to optimality under both. Lower, Equal, and Higher count negative, zero, and positive differences (\textsc{IR-IM-IS} minus \textsc{IR}), respectively.}
\label{tab:app-stress-quality}
\begin{tabular*}{0.86\textwidth}{@{\extracolsep{\fill}}lrlrrr@{}}
\toprule
Data set & Instances & Measure & Lower & Equal & Higher \\
\midrule
Stress & 47 & $\Delta_I$ & 0 & 47 & 0 \\
 & & $I_{\max}$ & 4 & 43 & 0 \\
 & & $I_\Sigma$ & 22 & 25 & 0 \\
\addlinespace
Stress-high & 21 & $\Delta_I$ & 0 & 21 & 0 \\
 & & $I_{\max}$ & 0 & 21 & 0 \\
 & & $I_\Sigma$ & 20 & 1 & 0 \\
\bottomrule
\end{tabular*}
\Description{Paired changes from selected IR to IR-IM-IS schedules on 47 Stress and 21 Stress-high instances solved to optimality under both. For each idle measure, the lower, equal, and higher counts sum to the instance count for that data set.}
\end{table}

\FloatBarrier
\subsection{Additional Encoding Results}
\label{app:extended-formula-profiling}

Table~\ref{tab:app-objective-layer-ablation} compares each alternative with
dense thresholds for the same objective. Direct idle-slot clauses give
the largest median per-instance clause reduction, for \textsc{IS}; attainable-value
compression yields smaller reductions for the other objectives.

Table~\ref{tab:app-formula-summary} compares median numbers of variables, clauses,
and literal occurrences on All-126. Relative to \textsc{Direct},
\textsc{Compact-Objective} reduced the median variable count most under
\textsc{BG-d2}, and had lower median clause and literal counts under all
three objectives.

\begin{table}[!htbp]
\centering
\caption{Complete encoding sizes on All-126 (126 instances). Counts and per-instance reductions relative to DT are medians; bold counts mark column minima within each objective, including rounding ties.}
\label{tab:app-objective-layer-ablation}
\begin{tabular*}{\textwidth}{@{\extracolsep{\fill}}llrrr@{\hspace{1em}}r@{}}
\toprule
Objective & Representation & Variables & Clauses & Literals & \shortstack{Clause reduction\\vs DT (\%)} \\
\midrule
\textsc{IS} & DT & 27{,}735 & 67{,}220 & 156{,}300 & -- \\
 & DI & \textbf{27{,}088} & \textbf{64{,}383} & \textbf{149{,}982} & 5.92 \\
\addlinespace[1pt]
\textsc{ISQ} & DT & 27{,}735 & 67{,}220 & 156{,}300 & -- \\
 & AV & \textbf{27{,}733} & \textbf{67{,}201.5} & \textbf{156{,}258.5} & 0.08 \\
\addlinespace[1pt]
\textsc{IM-IS} & DT & 27{,}742.5 & 67{,}353 & 156{,}669 & -- \\
 & AV & \textbf{27{,}739} & \textbf{67{,}318.5} & \textbf{156{,}581.5} & 0.15 \\
\addlinespace[1pt]
\textsc{IR-IS} & DT & 27{,}750 & 67{,}373.5 & 156{,}710 & -- \\
 & AV & \textbf{27{,}739} & \textbf{67{,}318.5} & \textbf{156{,}581.5} & 0.20 \\
\bottomrule
\end{tabular*}
\par\noindent\parbox{\textwidth}{\textit{Note:} DT, dense thresholds; DI, direct idle-slot clauses; AV, attainable-value compression.}
\Description{Eight rows compare dense thresholds with direct idle-slot clauses for IS or attainable-value compression for ISQ, IM-IS and IR-IS on 126 instances. Each row names its representation and gives median variables, clauses, literals and the paired median percentage clause reduction against dense thresholds for the same objective.}
\end{table}

\begin{table}[!htbp]
\centering
\caption{Median encoding sizes on All-126 (126 instances). Bold numerical values mark the smallest displayed count in each encoding pair, including rounding ties.}
\label{tab:app-formula-summary}
\begingroup
\setlength{\tabcolsep}{7pt}
\begin{tabular}{@{}llrrr@{}}
\toprule
Objective & Encoding & Variables & Clauses & Literals \\
\midrule
\textsc{BG-d2} & \textsc{Direct} & 79{,}877 & 170{,}517 & 424{,}056 \\
 & \textsc{Compact-Objective} \textbf{(ours)} & \textbf{27{,}736} & \textbf{66{,}318} & \textbf{153{,}675} \\
\addlinespace[1pt]
\textsc{IR} & \textsc{Direct} & 27{,}859 & 93{,}183.5 & 213{,}157.5 \\
 & \textsc{Compact-Objective} \textbf{(ours)} & \textbf{27{,}750} & \textbf{67{,}255.5} & \textbf{156{,}592} \\
\addlinespace[1pt]
\textsc{IR-IM-IS} & \textsc{Direct} & 27{,}859 & 93{,}609.5 & 213{,}577.5 \\
 & \textsc{Compact-Objective} \textbf{(ours)} & \textbf{27{,}750} & \textbf{67{,}381} & \textbf{156{,}717.5} \\
\bottomrule
\end{tabular}
\endgroup
\Description{Median formula counts for Direct and Compact-Objective under BG-d2, IR and IR-IM-IS on all 126 benchmark instances. Compact-Objective has the lower variable, clause and literal counts in each displayed pair; variable reductions are concentrated in BG-d2.}
\end{table}

\FloatBarrier

\section{Reproducibility Checklist for JAIR}
\label{app:reproducibility-checklist}

The answers refer to the manuscript and the repository described in Section~\ref{sec:code-data-availability}.

\subsection*{All articles}
\begin{enumerate}
  \item All claims investigated in this work are clearly stated.
  [\textbf{yes}]
  \item Clear explanations are given how the work reported substantiates the
  claims. [\textbf{yes}]
  \item Limitations or technical assumptions are stated clearly and explicitly.
  [\textbf{yes}]
  \item Conceptual outlines and/or pseudo-code descriptions of the AI methods
  introduced in this work are provided, and important implementation details
  are discussed. [\textbf{yes}]
  \item Motivation is provided for all design choices, including algorithms,
  implementation choices, parameters, data sets and experimental protocols
  beyond metrics. [\textbf{partially}]
\end{enumerate}

\subsection*{Articles containing theoretical contributions}
Does this paper make theoretical contributions? [\textbf{yes}] If yes, please complete the list below.
\begin{enumerate}
  \item All assumptions and restrictions are stated clearly and formally.
  [\textbf{yes}]
  \item All novel claims are stated formally (e.g., in theorem statements).
  [\textbf{yes}]
  \item Proofs of all non-trivial claims are provided in sufficient detail to
  permit verification by readers with a reasonable degree of expertise (e.g.,
  that expected from a PhD candidate in the same area of AI). [\textbf{yes}]
  \item Complex formalism, such as definitions or proofs, is motivated and
  explained clearly. [\textbf{yes}]
  \item The use of mathematical notation and formalism serves the purpose of
  enhancing clarity and precision; gratuitous use of mathematical formalism
  (i.e., use that does not enhance clarity or precision) is avoided.
  [\textbf{yes}]
  \item Appropriate citations are given for all non-trivial theoretical tools
  and techniques. [\textbf{yes}]
\end{enumerate}

\subsection*{Articles reporting on computational experiments}
Does this paper include computational experiments? [\textbf{yes}] If yes, please complete the list below.
\begin{enumerate}
  \item All source code required for conducting experiments is included in an
  online appendix or will be made publicly available upon publication of the
  paper. The online appendix follows best practices for source code readability
  and documentation as well as for long-term accessibility.
  [\textbf{partially}]
  \item The source code comes with a license that allows free usage for
  reproducibility purposes. [\textbf{yes}]
  \item The source code comes with a license that allows free usage for research
  purposes in general. [\textbf{yes}]
  \item Raw, unaggregated data from all experiments is included in an online
  appendix or will be made publicly available upon publication of the paper.
  The online appendix follows best practices for long-term accessibility.
  [\textbf{partially}]
  \item The unaggregated data comes with a license that allows free usage for
  reproducibility purposes. [\textbf{yes}]
  \item The unaggregated data comes with a license that allows free usage for
  research purposes in general. [\textbf{yes}]
  \item If an algorithm depends on randomness, then the method used for
  generating random numbers and for setting seeds is described in a way
  sufficient to allow replication of results. [\textbf{partially}]
  \item The execution environment for experiments, the computing infrastructure
  (hardware and software) used for running them, is described, including GPU/CPU
  makes and models; amount of memory (cache and RAM); make and version of
  operating system; names and versions of relevant software libraries and
  frameworks. [\textbf{partially}]
  \item The evaluation metrics used in experiments are clearly explained and
  their choice is explicitly motivated. [\textbf{yes}]
  \item The number of algorithm runs used to compute each result is reported.
  [\textbf{yes}]
  \item Reported results have not been ``cherry-picked'' by silently ignoring
  unsuccessful or unsatisfactory experiments. [\textbf{yes}]
  \item Analysis of results goes beyond single-dimensional summaries of
  performance (e.g., average, median) to include measures of variation,
  confidence, or other distributional information. [\textbf{yes}]
  \item All (hyper-)parameter settings for the algorithms/methods used in
  experiments have been reported, along with the rationale or method for
  determining them. [\textbf{partially}]
  \item The number and range of (hyper-)parameter settings explored prior to
  conducting final experiments have been indicated, along with the effort spent
  on (hyper-)parameter optimisation. [\textbf{NA}]
  \item Appropriately chosen statistical hypothesis tests are used to establish
  statistical significance in the presence of noise effects. [\textbf{NA}]
\end{enumerate}

\subsection*{Articles using data sets}
Does this work rely on one or more data sets (possibly obtained from a benchmark
generator or similar software artifact)? [\textbf{yes}] If yes, please complete the list below.
\begin{enumerate}
  \item All newly introduced data sets are included in an online appendix or
  will be made publicly available upon publication of the paper. The online
  appendix follows best practices for long-term accessibility with a license
  that allows free usage for research purposes. [\textbf{partially}]
  \item The newly introduced data set comes with a license that allows free
  usage for reproducibility purposes. [\textbf{partially}]
  \item The newly introduced data set comes with a license that allows free
  usage for research purposes in general. [\textbf{partially}]
  \item All data sets drawn from the literature or other public sources
  (potentially including authors' own previously published work) are accompanied
  by appropriate citations. [\textbf{yes}]
  \item All data sets drawn from the existing literature (potentially including
  authors' own previously published work) are publicly available.
  [\textbf{yes}]
  \item All new data sets and data sets that are not publicly available are
  described in detail, including relevant statistics, the data collection
  process and annotation process if relevant. [\textbf{yes}]
  \item All methods used for preprocessing, augmenting, batching or splitting
  data sets (e.g., in the context of hold-out or cross-validation) are described
  in detail. [\textbf{yes}]
\end{enumerate}

\subsection*{Explanations on the answers above}
The repository provides source code, run results, and instructions for reproducing the analyses. It documents instance generation, random seeds, and experimental settings. Availability and licensing are described in Section~\ref{sec:code-data-availability}.

Partial responses concern permanent archiving, documentation of solver seeds and execution environments, and the rationale for some fixed parameter choices. Parameter values were fixed before evaluation, without parameter search. The reported comparisons are descriptive.

\end{document}